%% file: main.tex
\documentclass{article}
\usepackage{spconf}
\usepackage{amsmath,graphicx}
\usepackage{amssymb,amsmath,bm}
\usepackage{textcomp}
\usepackage{algorithm}
\usepackage{algpseudocode}
\usepackage{float}
\usepackage{multirow}
\usepackage{mathtools}
\usepackage{placeins}
\usepackage{subcaption}  
\usepackage[justification=centering]{caption}

\graphicspath{{../figures/}}

\usepackage{tikz,placeins}
\usetikzlibrary{arrows,decorations.markings}
\tikzstyle{every picture}+=[font=\rmfamily\it\bfseries\large]
\usetikzlibrary{positioning, shapes}
\pgfdeclarelayer{background}
\pgfdeclarelayer{foreground}
\pgfsetlayers{background,main,foreground}

\newcommand{\specialcell}[2][c]{%
     \begin{tabular}[#1]{@{}c@{}}#2\end{tabular}}

\title{Small energy masking for improved neural network training for 
end-to-end speech recognition}
\name{Chanwoo Kim, Kwangyoun Kim, and Sathish Reddy Indurthi
  \thanks{Thanks to Samsung Electronics for funding this research. 
The authors are thankful to Executive Vice President Seunghwan Cho 
and speech processing Lab. members at Samsung Research.}}

\address{Samsung Research, Seoul, South Korea \\
  {\small \tt \{chanw.com, ky85.kim, s.indurthi\}@samsung.com }}
\begin{document}
\ninept
\maketitle
\begin{abstract}
  In this paper, we present a Small Energy Masking (SEM) algorithm, which
masks inputs having values below a certain threshold. 
  More specifically, a time-frequency bin is masked if the filterbank energy 
in this bin is less than a certain energy threshold.
 A uniform distribution is employed to randomly generate the ratio of 
 this energy threshold to the peak filterbank energy of each utterance in decibels.
The unmasked feature elements are scaled so that 
  the total sum of the feature values remain the same through this masking
  procedure.
This very simple algorithm shows relatively 11.2 \%
and 13.5 \% Word Error Rate (WER) improvements on the standard
  LibriSpeech {\tt test-clean} and {\tt test-other } sets over 
  the baseline end-to-end speech recognition system. 
  Additionally, compared to the input dropout 
algorithm, SEM algorithm shows relatively 7.7 \% and 11.6 \% 
  improvements on the same  LibriSpeech {\tt test-clean} and {\tt test-other} sets.
With a modified shallow-fusion technique with a Transformer LM,
we obtained a 2.62 \% WER on the LibriSpeech {\tt test-clean} set and a 7.87 \% WER
on the LibriSpeech {\tt test-other} set.
\end{abstract}
\section{Introduction}
\label{sec:intro}
Recently, deep learning techniques have significantly 
improved speech recognition accuracy \cite{
G_Hinton_IEEE_Signal_Process_Mag_2012}.
These improvements have been obtained by the shift from Gaussian Mixture Model
(GMM) to the Feed-Forward Deep Neural Networks (FF-DNNs), FF-DNNs
to Recurrent Neural Network (RNN) such as the Long Short-Term Memory
(LSTM) networks \cite{S_Hochreiter_neural_computation_1997_00}. 
Thanks to these advances, voice assistant devices such as Google Home
\cite{c_kim_interspeech_2017_00}
, Amazon Alexa and Samsung Bixby \cite{samsung_bixby} are widely used at  
home environments.

Recently there has been tremendous amount of research in switching
from the conventional Weighted Finite State Transducer (WFST)
based decoder using an Acoustic Model (AM) and a Language Model (LM)
to a complete end-to-end all-neural speech recognition systems 
\cite{w_chan_icassp_2016_00,  j_chorowski_nips_2015_00}. 
These complete end-to-end systems have started surpassing the performance of
the conventional WFST-based decoders with a very large training 
dataset \cite{c_chiu_icassp_2018_00}, a better choice of target unit 
such as Byte Pair Encoded (BPE) subword units, and an improved 
training methodology such as Minimum
Word Error Rate (MWER) training \cite{r_Prabhavalkar_icassp_2018_00}.

Overfitting has been a major issue in training a large size neural network
model for a long time.
The dropout approach \cite{n_srivastava_jmlr_2014_00} has been applied
to overcome this issue in which both the input and the hidden 
units are randomly dropped out to regularize the network.
In the case of the input dropout, the input feature elements are masked 
with a certain fixed probability of $r$, and the 
remaining input feature elements are scaled up by $1.0 / (1.0 - r)$.

One open question regarding the input dropout is whether it is always
a good idea to dropout input feature elements completely randomly.
In our previous work 
\cite{C_Kim_ASRU_2009_2}, it has been observed that feature elements in 
time-frequency bins with smaller power are more adversely affected by additive noise.
Motivated by this observation, we intentionally boosted power or energy in such
time-frequency bins in  \cite{C_Kim_ASRU_2009_2}. This algorithm is 
referred to as the Small Power Boosting (SPB) algorithm. This approach proved to be 
quite successful for very difficult noisy environments such as corruption
by background music.
\begin{figure}[tbp]
  \begin{subfigure}{0.5\textwidth}
    \centering
    \includegraphics[width=60mm]{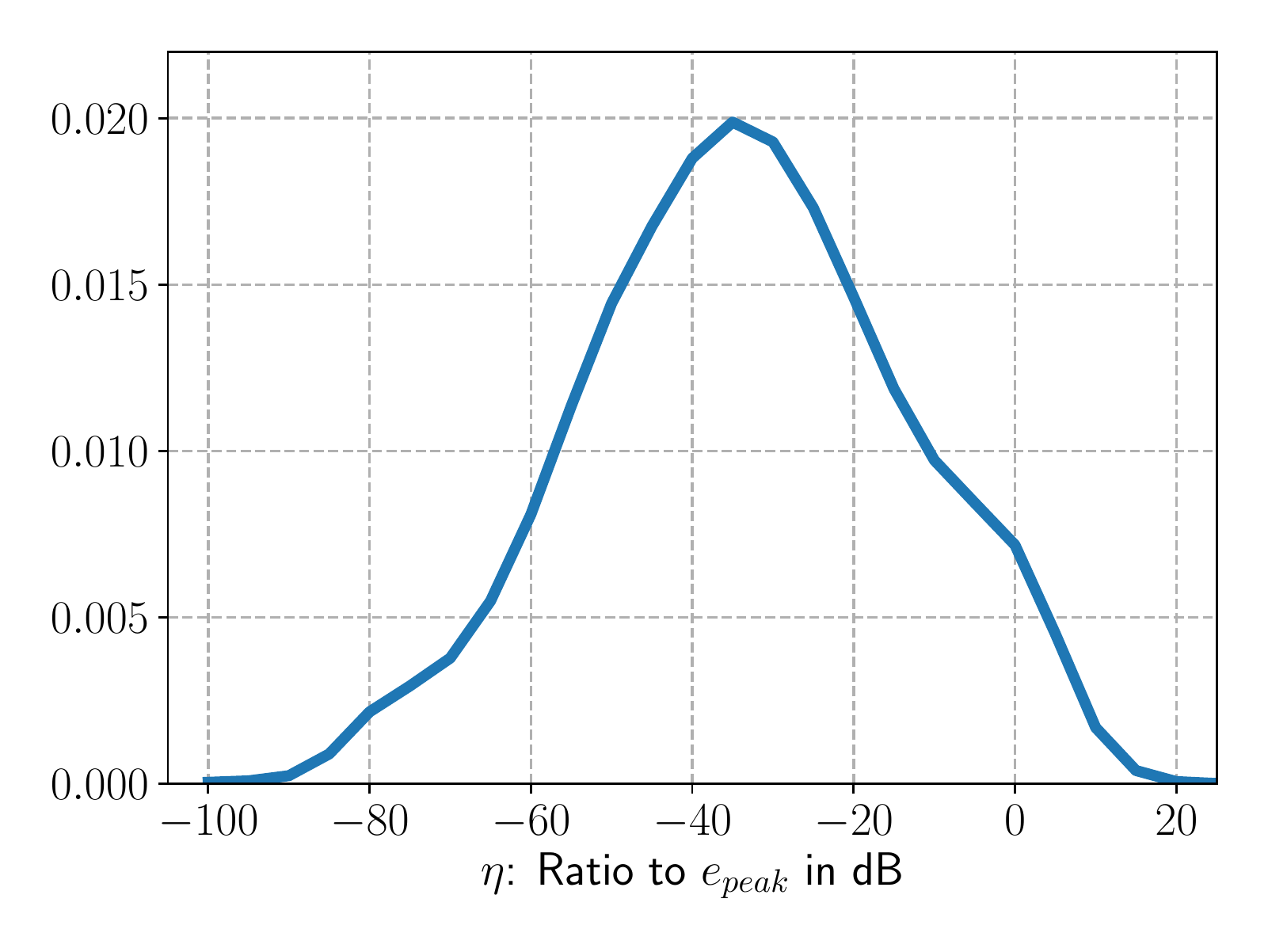}
    \caption {
      \label{fig:distribution_of_filterbank_energy}
    }
  \end{subfigure} \\
  \begin{subfigure}{0.5\textwidth}
    \centering
    \includegraphics[width=60mm]{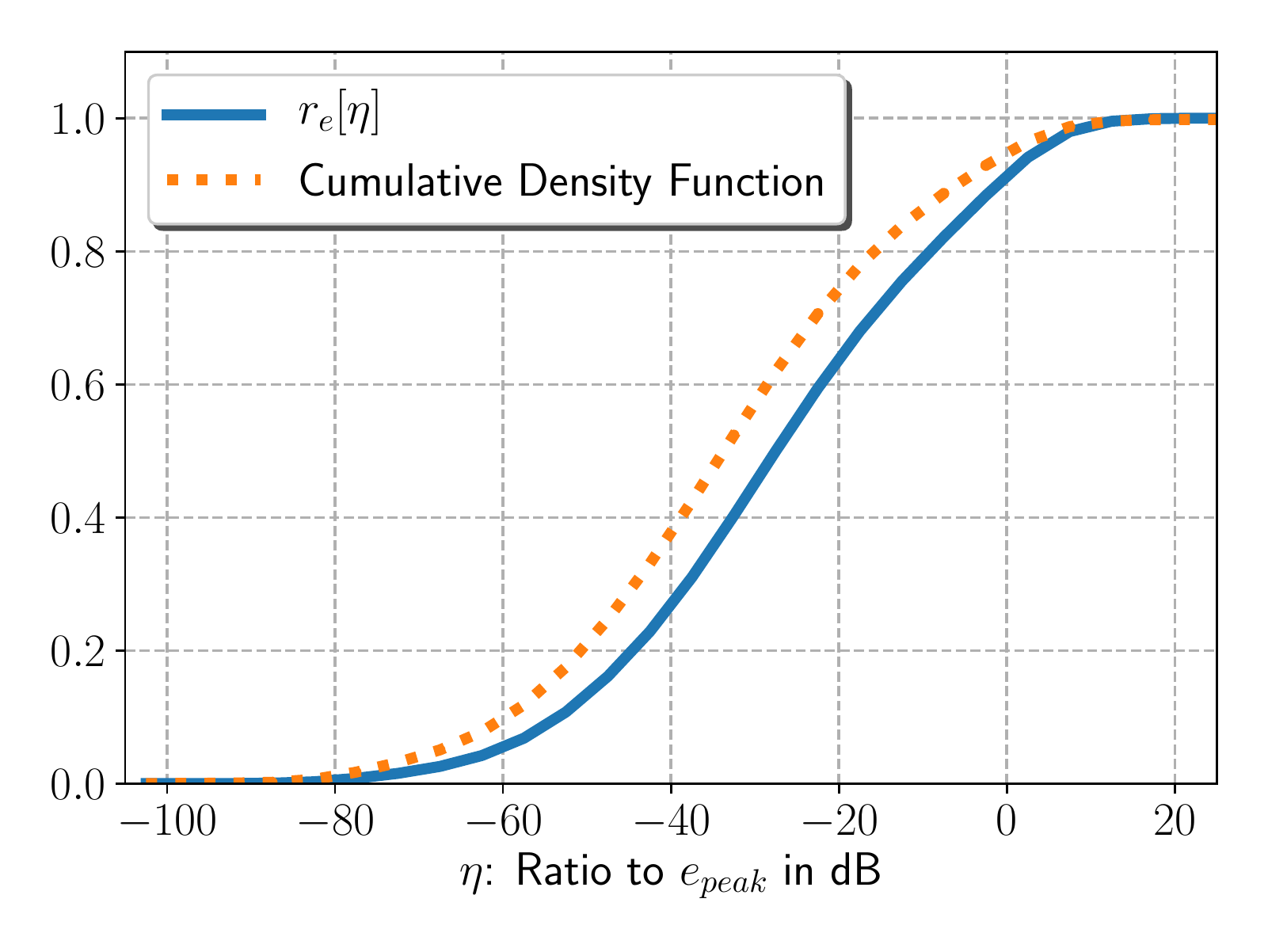}
    \caption{
      \label{fig:cdf_masked_portion}
    }
  \end{subfigure} 
  \caption{(a) The probability density of $\eta$ in \eqref{eq:def_eta}, 
  which is the relative ratio to $e_{\text{peak}}$ in dB.
  (b) The cumulative density of the same $\eta$  and
  the portion of the filterbank energy below $\eta$, which is defined
  as $r_{e}(\eta_{th})$ in \eqref{eq:ratio_energy}.
  \vspace{-5mm}
  }
\end{figure}
In this paper, we present an algorithm referred to as {\it Small Energy
Masking} (SEM). Unlike the input dropout approach, we mask time-frequency
bins in the spectral domain whose filterbank energy is less than a certain 
energy threshold. This energy threshold is chosen from a random uniform 
distribution. This masking is applied to the input features. For the 
unmasked components of each feature, we scale these values so that the total sum
of the feature values remain the same. This algorithm is also 
different from SpecAugment in \cite{s_park_interspeech_2019_00} in that
masking is done for time-frequency bins with smaller energies.
\section{Distribution of energy in time-frequency bins of speech signals}
\label{sec:format}
Before describing the Small Energy Masking (SEM) algorithm 
in detail in Sec. \ref{sec:small_energy_masking_algorithm}, 
we first look into the distribution of energy in each time-frequency 
bin. %
\begin{figure}[tbp]
  \begin{center}
    \resizebox{80mm}{!}{\input{sem_procedure_diagram}}
     \caption {
      \label{fig:sem_procedure_diagram}
       The procedure of applying the Small Energy Masking (SEM).
     }
  \end{center}
\vspace{-5mm}
\end{figure}
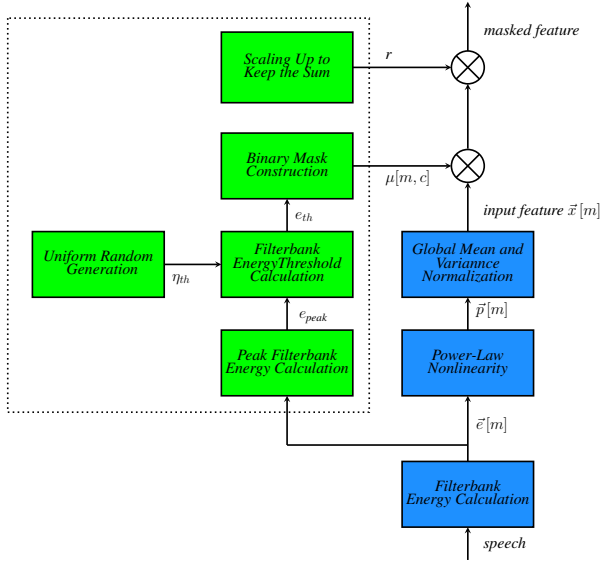
The filterbank energy $e[m, c]$ in each time-frequency bin is calculated 
using the following equation \cite{c_kim_asru_2019_00}:
\begin{align}
  e[m, c] = \sum_{k=0}^{K/2} |X[m, e^{j \omega_k} ] |^2  M_c[e^{j\omega_k}] 
  \label{eq:filterbank_energy}
\end{align}
were $m$ is the frame index, $c$ is the filterbank channel index, $K$
is the Fast Fourier Transform (FFT) size. $\omega_k$ is the discrete-time
frequency  defined by $\omega_k = \frac{2 \pi k}{K}$ where $0 \le k \le K-1$, 
and $M_c[e^{j\omega_k}]$ is the triangular mel response for the $l$-th 
channel \cite{pmermelstein1975}. Throughout this paper, in obtaining 
the spectrum $X[m, e^{j \omega_k}]$ in \eqref{eq:filterbank_energy}, 
we use Hamming windows of the 25 {\it ms} length and 
the 10 {\it ms} period between successive frames.
As in \cite{C_Kim_ASRU_2009_2}, we define the  peak filterbank energy 
$e_{\text{peak}}$ of each utterance as follows:
\begin{align}
  e_{\text{peak}} \coloneqq \text{The 95-th percentile of\;} e[m, c] 
  \text{\;in \eqref{eq:filterbank_energy} for an utterance.}
  \label{eq:def_e_peak}
\end{align}
Since $e_{\text{peak}}$ is calculated within a single utterance,
the actual value may be different for different utterances.
For a certain filterbank energy value $e[m, c]$, let us define the
following term $\eta$ and the function $f$ that is the ratio
of $e[m, c]$ to $e_\text{peak}$ in decibels (dB):
\begin{align}
  \eta = f\left(e[m, c]\right) \coloneqq 10 \log_{10} 
      \left( \frac{e[m, c]}{e_{\text{peak}}}  \right).
  \label{eq:def_eta}
\end{align}

Fig. \ref{fig:distribution_of_filterbank_energy} shows the probability
density function of the time-frequency bins with respect to
$\eta$ defined in \eqref{eq:def_eta}.  In obtaining this distribution, 
we used a randomly chosen 1,000 utterances 
from the LibriSpeech training corpus \cite{v_panayotov_icassp_2015_00}.
As shown in Fig. \ref{fig:distribution_of_filterbank_energy},
the filter bank energy in each time-frequency bin is mostly distributed
between -90 dB and 10 dB
with respect to $e_{\text{peak}}$.  Fig. \ref{fig:cdf_masked_portion} 
shows the cumulative density function of $\eta$ in \eqref{eq:def_eta}
and the following function:
\begin{align}
  r_{e}(\eta_{th}) = \frac{\sum_{f\left( e[m,c] \right) < \eta_{th}} e[m,
  c]}{\sum e[m, c]},
  \label{eq:ratio_energy}
\end{align}
which is the ratio of the energy when $\eta_{th}$ is given as the ratio
threshold with respect to $e_{\text{peak}}$.
\begin{figure}
  \centering
  \begin{subfigure}{1.0\linewidth}
    \centering\includegraphics[width=0.8\linewidth]{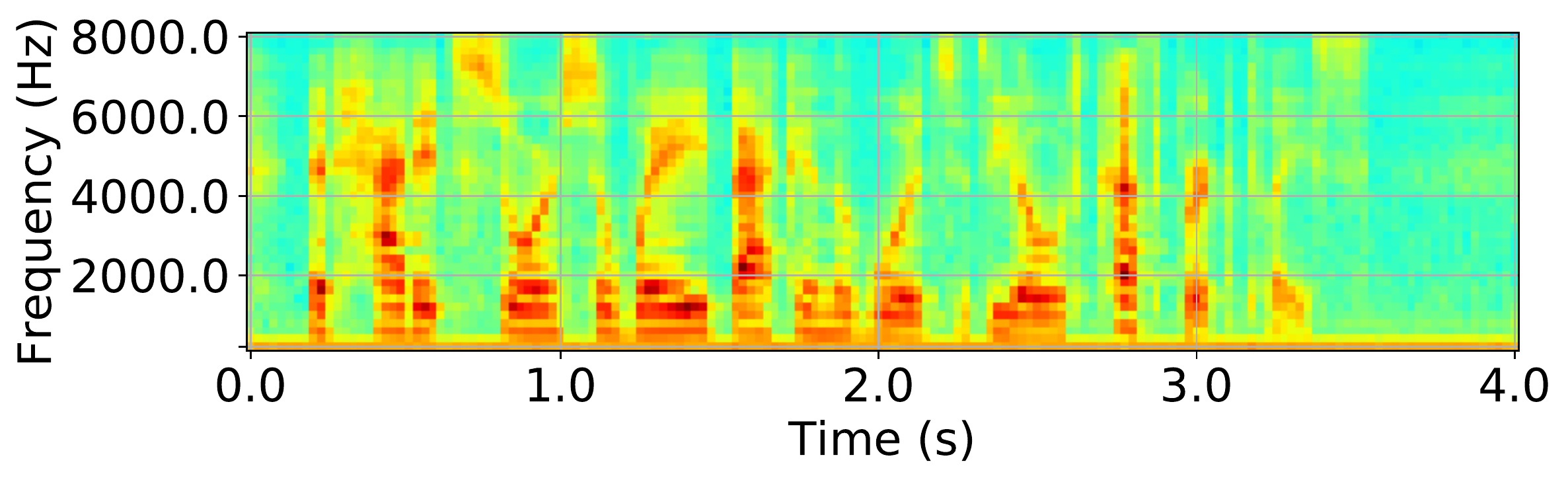}
    \caption{
      \label{fig:fig_spectrogram_clean}
    }
  \end{subfigure} \\
  \begin{subfigure}{1.0\linewidth}
    \centering\includegraphics[width=0.8\linewidth]{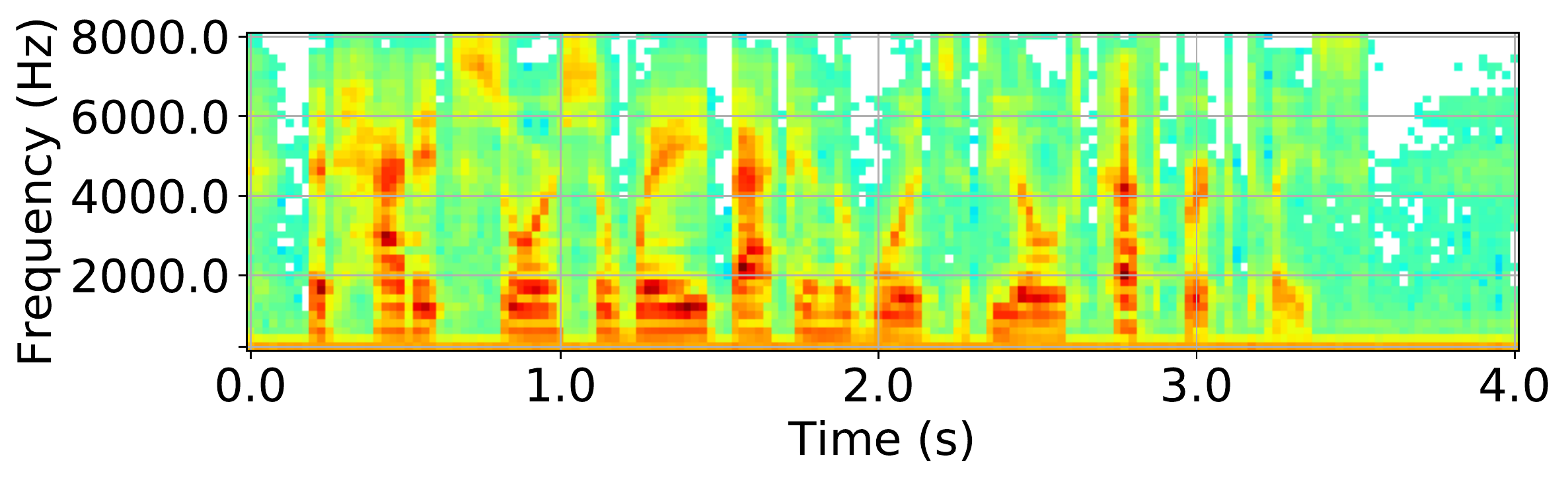}
    \caption{
      \label{fig:fig_spectrogram_m40db}
    }
  \end{subfigure} \\
  \begin{subfigure}{1.0\linewidth}
    \centering\includegraphics[width=0.8\linewidth]{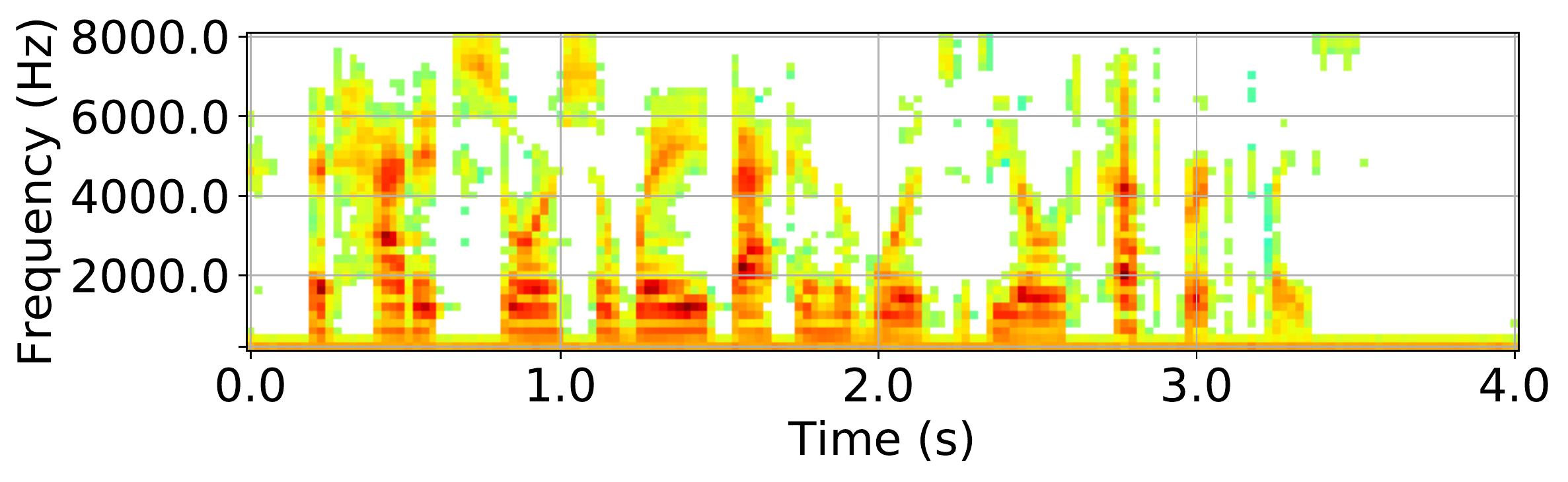}
    \caption{
      \label{fig:fig_spectrogram_m20db}
    }
  \end{subfigure} \\
  \begin{subfigure}{1.0\linewidth}
    \centering\includegraphics[width=0.8\linewidth]{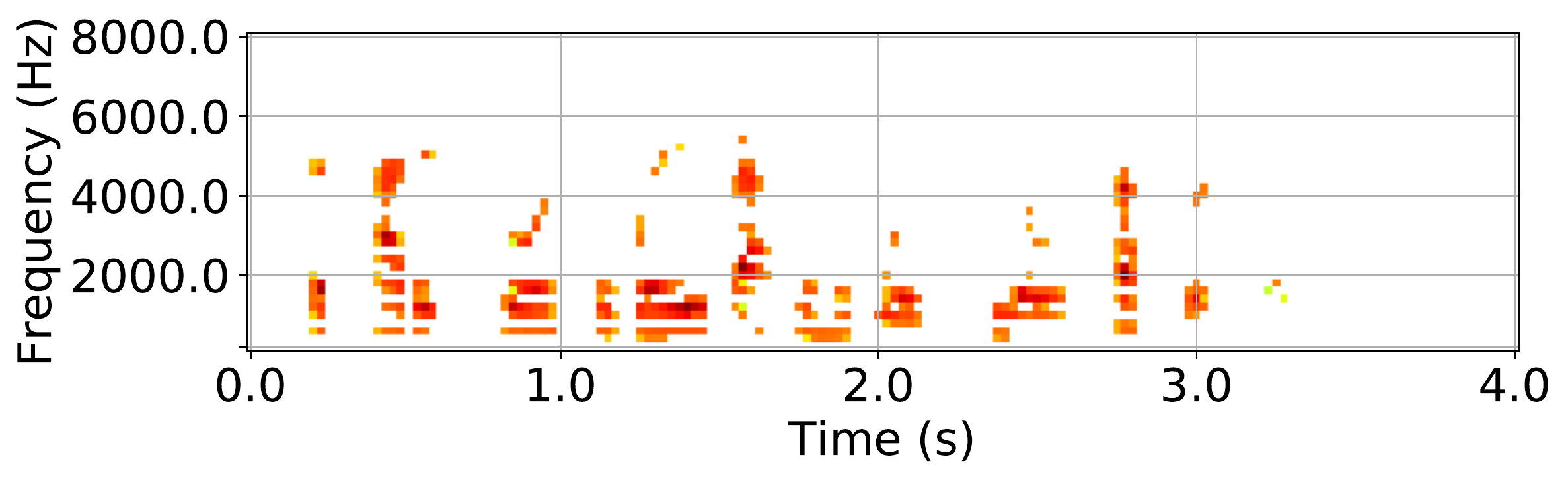}
    \caption{
      \label{fig:fig_spectrogram_0db}
    }
  \end{subfigure}
  \caption{(a) The power-mel spectrogram of the original speech.
  Small Energy Masked power-mel spectrograms with different 
  $\eta_{th}$ values:
  (b) $\eta_{th} = -40$ dB, (c) $\eta_{th} = -20$ dB, and 
  (d) $\eta_{th} = 0$ dB.
  \label{fig:spectrogram}
  }

\end{figure}
\section{Small Energy Masking algorithm}
\label{sec:small_energy_masking_algorithm}
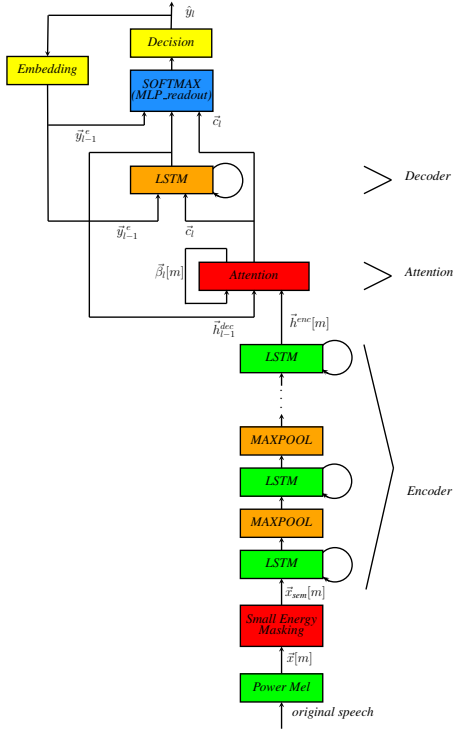
\begin{figure}[tbp]
  \begin{center}
    \resizebox{60mm}{!}{\input{e2e_asr_block_diagram}}
      \caption {  
      The structure of the entire end-to-end
      speech recognition system.
     }
     \label{fig:e2e_asr_block_diagram}
  \end{center}
\vspace{-5mm}
\end{figure}
The procedure of applying the Small Energy Masking (SEM) to the input
feature is shown in Fig. \ref{fig:sem_procedure_diagram}. 
More specifically, procedures related to masking is depicted inside 
the dotted rectangle.
As shown in this figure, we first obtain the filterbank energy $e[m, c]$
in each time frequency bin using \eqref{eq:filterbank_energy}.
The mel filterbank energy vector $\vec{e}\,[m]$ for the frame index $m$ 
is defined from the following equation:
\begin{align}
  \vec{e}\,[m] = \Big[e[m, 0],\,e[m, 1],\,e[m, 2]\,\cdots\,e[m, C-1]\Big]^T
\end{align}
where $C$ is the number of filterbank channels.
For each utterance, we obtain the peak filterbank energy 
$e_{\text{peak}}$ using \eqref{eq:def_e_peak}.
We generate a random threshold $\eta_{th}$ that is a ratio to 
$e_{\text{peak}}$ in dB from the following uniform distribution:
\begin{align}
       \eta_{th} \sim \mathcal{U}(\eta_a, \eta_b).
    \label{eq:uniform_distribution}
\end{align}
where $\eta_a$ and $\eta_b$ are lower and upper bounds of this 
uniform distribution. 
In our experiments in Sec. \ref{sec:experimental_results}, we observe that 
$\eta_a$ = -80 and $\eta_b$ = 0 are appropriate.

Using the $\eta_{th}$, the filterbank energy threshold is obtained by the
following equation derived from \eqref{eq:def_eta}:
\begin{align}
  e_{th} = e_{\text{peak}} 10^{\frac{\eta_{th}}{10}}.
\end{align}
The binary mask is generated using the following rule:
\begin{align}
  \mu[m, c] =  
  \begin{cases}
    1, \, \qquad e[m, c] \ge e_{th},   \\
    0, \, \qquad e[m, c] < e_{th}.
  \end{cases}
\end{align}
The masked feature is generated in the following way:
\begin{align}
  x_{\mu}[m, c] = x[m, c] \mu[m, c] 
  \label{eq:apply_masking}
\end{align}
where $x[m, c]$ is an element of the power-mel feature, which is obtained 
by the power-law nonlinearity of $(\cdot)^{\frac{1}{15}}$ 
\cite{C_Kim_INTERSPEECH_2009_2, c_kim_taslp_2016_00} on the mel filterbank energy $e[m, c]$.

Fig. \ref{fig:spectrogram} shows the masked power-mel spectrogram with 
different values of the threshold $\eta_{th}$.  For example, Fig. 
\ref{fig:fig_spectrogram_m20db} shows the power-mel spectrogram when
$\eta_{th} = $- 20 dB. Compared to the original power-mel spectrogram
in \ref{fig:fig_spectrogram_clean}, we may observe that approximately
70 \% of the time-frequency bins in Fig. \ref{fig:fig_spectrogram_m20db} are 
masked. From Fig. \ref{fig:cdf_masked_portion}, we observe 
that when $\eta_{th} = $ -20dB, on average, 74.3 \% of time-frequency bins   
are masked.
As shown in Fig. \ref{fig:sem_procedure_diagram}, we apply the ``global"
mean and variance normalization as in
\cite{a_zeyer_interspeech_2018_00}, since the 
utterance-by-utterance mean and variance 
normalizations are not easily realizable
for streaming speech recognition \cite{k_kim_asru_2019_00}. Note that 
 mean subtraction must be applied before masking, otherwise, the 
non-zero values in the masked region will distort the model during
the training.
Through the masking procedure in \eqref{eq:apply_masking}, the 
sum of feature elements for each utterance is kept the same using the 
following scaling coefficient:
\begin{align}
  r = \frac{\sum_{\text{for each utt.}} x[m, c]}
    {\sum_{\text{for each utt.}} x_{\mu}[m, c]},
\end{align}
The final input feature to the neural network $x_{\text{sem}}[m, c]$ is 
given by:
\begin{align}
  x_{\text{sem}}[m, c] = r x_{\mu}[m, c]. 
    \label{eq:sem_element}
\end{align}
\section{End-to-end Speech Recognition System}
\begin{table}[!htbp]
  \renewcommand{\arraystretch}{1.3}
  \centering
  \caption{\label{tbl:sem_lower_bound_fixed}
    Word Error Rates (WERs) obtained on the LibriSpeech test set
    \cite{v_panayotov_icassp_2015_00}  using the SEM algorithm.
    In this case, in \eqref{eq:uniform_distribution},  the $\eta_a$ 
    value is fixed at -80 dB and $\eta_b$ values is changed.
  }
  \begin{tabular}{| c || c | c | c| c | c | }
			\hline
                             $  \eta_{b} $
                             & -60 dB   
                             & -40 dB    
                             & -20 dB    
                             &   0 dB  
                             &  \textcolor{blue}{\textbf{baseline}} \\
			\hline \hline
							 test-clean  &     4.03  \%  &  4.05 \%  &  3.89 \% & \textcolor{blue}{\textbf{3.72}} \%  & 4.19 \% \\  
               test-other  &   13.64 \% &  13.69  \% &   12.74 \%  &  \textcolor{blue}{\textbf{11.65}} \%  & 13.47 \%   \\  
               average     &   8.84  \%  & 8.87 \%  &  8.32 \%  & \textcolor{blue}{\textbf{7.69}}
               \%  & 8.83 \% \\
			\hline
  \end{tabular}
\end{table}
\begin{table}[!htbp]
  \renewcommand{\arraystretch}{1.3}
  \centering
  \caption{\label{tbl:sem_upper_bound_fixed}
    Word Error Rates (WERs) obtained on the LibriSpeech test set
    \cite{v_panayotov_icassp_2015_00}  using the SEM algorithm. 
    In this case, in \eqref{eq:uniform_distribution}, the $\eta_a$ value 
    is changed and $\eta_b$ is fixed at 0 dB.
  }
  \begin{tabular}{| c ||  c | c | c | c | c |}
			\hline
                             $ \eta_a  $
                             & -20 dB
                             & -40 dB 
                             & -60 dB  
                             & -80 dB 
                             & \textcolor{blue}{\textbf{baseline}} \\
			\hline \hline
               test-clean  &  45.15 \% &   6.57 \% &  4.07 \% & \textcolor{blue}{\textbf{3.72}} \% & 4.19 \%  \\
               test-other   & 77.71 \%  &  20.43 \% & 12.73 \% & \textcolor{blue}{\textbf{11.65}} \%
                  & 13.47 \% \\


               average      & 61.43 \% &  13.5 \% &  8.40 \% &  
               \textcolor{blue}{\textbf{7.69}}  \% & 8.83 \%  \\
			\hline
  \end{tabular}
\end{table}
\begin{table}[!htbp]
  \renewcommand{\arraystretch}{1.3}
  \centering
  \caption{ \label{tbl:fixed_threshold}
    Word Error Rates (WERs) obtained on the LibriSpeech corpus 
    \cite{v_panayotov_icassp_2015_00} using a ``fixed" ratio threshold
    $\eta_{th}$  instead of the random threshold generated by the uniform distribution in \eqref{eq:uniform_distribution}.
  }
  \begin{tabular}{| c || c | c | c | c | c |}
			\hline
                             $ \eta_{th} $
                             & \specialcell{\textcolor{blue}{\textbf {baseline} } 
                                    \\ $-\infty$ dB} 
                             & -80 dB
                             & -70 dB
                             & -60 dB
                             & -50 dB  \\
			\hline \hline
               test-clean  & \textcolor{blue}{\textbf{ 4.19 }} \% & 4.27 \%
               & 4.26 \% & 4.31 \% & 
               4.52 \% \\ 
               test-other  &  \textcolor{blue}{\textbf{ 13.47 }} \% & 13.92 \%
               & 13.93 \% & 14.09 \% &
               15.67 \% \\
               average & \textcolor{blue}{\textbf{8.83 }} \% & 9.10 \% & 9.10
               \% & 9.20 \% & 10.10 \%  \\ 
			\hline
  \end{tabular}
\end{table}
\begin{table}[!htbp]
  \renewcommand{\arraystretch}{1.3}
  \centering
  \caption{\label{tbl:input_dropout}
    Word Error Rates (WERs) obtained on the LibriSpeech corpus 
    \cite{v_panayotov_icassp_2015_00} using the input dropout method in 
    \cite{n_srivastava_jmlr_2014_00}. \\ $r$ is the dropout rate.
  }
  \begin{tabular}{| c || c | c | c | c |}

			\hline
                             & \specialcell{  
                             \textcolor{blue}{\textbf {baseline} } \\ $r = 0$} 
                             & $r = 0.1$ 
                             & $r = 0.2$
														 & $r = 0.3$ \\
			\hline \hline
               test-clean  &    4.19  \% & \textcolor{blue}{\textbf{ 4.03 }} \%
               &   4.29 \%  &  4.27 \%  \\  
               test-other  &  13.47  \% & \textcolor{blue}{\textbf{13.18}} \%
               &  13.77 \%  & 14.59 \%  \\  
							 average     &  8.83  \% & \textcolor{blue}{\textbf{8.61}}  \%
               & 9.03  \%  & 9.43  \%  \\  
			\hline
  \end{tabular}
\end{table}
The structure of our entire end-to-end speech recognition system
 is shown in Fig.~\ref{fig:e2e_asr_block_diagram}.
Our speech recognition system is a modified version of the previous one 
introduced in \cite{c_kim_asru_2019_01} with the architecture
and the pre-training schemes motivated by \cite{a_zeyer_interspeech_2018_00}.
$\vec{x}[m]$ and $\vec{y}_l$ are the input power mel 
filterbank vector and the output label,
respectively.
$m$ is the input frame index and $l$ is the decoder output
step index. 
We use the power mel filterbank vector instead of the more 
widely used log mel filterbank vector based on our previous result 
\cite{c_kim_interspeech_2019_00}.
$\vec{x}_{\text{sem}}[m]$ is the Small Energy Masked (SEM)
feature vector from \eqref{eq:sem_element} and defined as follows:
\begin{align}
  \vec{x}_{\text{sem}}[m] = \Big[x_{\text{sem}}[m, 0],\,x_{\text{sem}}[m,
  1],\,x_{\text{sem}}[m, 2]\,\cdots\,x_{\text{sem}}[m, C-1]\Big]^T
\end{align}
where $C$ is the number of filterbank channels.
$\vec{c}_l$ is the context vector calculated 
as a weighted sum of the encoder hidden state vectors denoted as $\vec{h}^{enc}[m]$.
The attention weights are computed as a softmax of energies computed as  
a function of the encoder hidden state $\vec{h}^{enc}[m]$, the decoder hidden state
$\vec{h}^{dec}_l$, and the attention weight
feedback $\vec{\beta}_l[m]$  \cite{a_zeyer_interspeech_2018_00}.

In \cite{a_zeyer_interspeech_2018_00}, the peak value of the speech 
waveform is normalized to be one. However, since finding the peak sample value 
is not possible for online feature extraction, we do not perform this
normalization. We modified the input pipeline
so that the online feature generation can be performed. We disabled the 
clipping of feature range between -3 and 3, which is the default 
setting for the LibriSpeech experiment in \cite{a_zeyer_interspeech_2018_00}. 
The encoder consists of six layers of bi-directional Long Short-Term
Memories (LSTMs) interleaved with 2:1 maxpool layers in the bottom three layers.
Thus, the entire time reduction factor is 8.
 The decoder is a single layer of uni-directional Long Short-Term Memory
 (LSTM). For all the LSTM layers, we used the cell size of 1024.
 For the better stability during the training, we use the gradient clipping by  
global norm \cite{r_pascanu_icml_2013}, which is implemented as  
{\tt tf.clip\_by\_global\_norm } API in Tensorflow  \cite{m_abadi_usenix_2016}.
\section{Experimental Results}
\label{sec:experimental_results}
In this section, we present speech recognition results 
obtained using the SEM algorithm on the LibriSpeech database 
\cite{v_panayotov_icassp_2015_00}.
For training, we used the entire 960 hours LbriSpeech training
set consisting of 281,241 utterances. For evaluation, we used the
official 5.4 hours {\tt test-clean} and 5.1 hours {\tt test-other} sets.
We conducted experiments
using the power mel filterbank vector of size of 40 as in 
\cite{c_kim_asru_2019_00, c_kim_interspeech_2019_00}.
We expect that the performance of SEM will depend on the selection
of the lower and upper bounds of the uniform distribution in 
\eqref{eq:uniform_distribution}. In Fig.
\ref{fig:distribution_of_filterbank_energy}, we observe that $\eta$ 
values in \eqref{eq:def_eta} are mostly concentrated between -80 dB and 0 dB.
Thus, we decide to choose the boundaries of the uniform distribution 
in this range. In the first set of the experiments, we fixed the lower bound
of this uniform distribution $\eta_a$ at -80 dB and changed the upper bound
$\eta_b$. These experimental results are shown in Table
\ref{tbl:sem_lower_bound_fixed}.
In this case, the best performance was obtained when $\eta_b$ = 0 dB.
In the second set of the experiments, we fixed the upper bound $\eta_b$ at 0 dB
and changed the lower bound $\eta_a$. This result is summarized in Table
\ref{tbl:sem_upper_bound_fixed}. As shown in this Table,
if $\eta_a$ value is too large, performance significantly degrades, 
which means that masking should not be too aggressive all the time. 
As shown in the
Tables \ref{tbl:sem_lower_bound_fixed} and \ref{tbl:sem_upper_bound_fixed},
we obtain 3.72 \% and 11.65 \% WERs on the LibriSpeech 
{\tt test-clean} and {\tt test-other} respectively, which are relatively
11.2 \% and 13.5 \% improvements, respectively.

In order to find out whether the randomization of the energy threshold 
in SEM is important, we repeated the same kind of experiments with a fixed 
energy threshold $\eta_{th}$. 
As shown in Table \ref{tbl:fixed_threshold}, it hardly shows 
any improvement over its baseline for the range of
values between -80 dB and -50 dB.
From this result, we conclude that
the randomization of the energy threshold plays a critical role
during the training.

Finally, to compare the performance of our SEM algorithm with the 
well known input dropout approach, we conducted the same set of
experiments using the input dropout algorithm. The result is summarized in
Table \ref{tbl:input_dropout}. Input dropout shows relatively 3.81 \%
and 2.15 \% improvements on the LibriSpeech {\tt test-clean} and {\tt test-other} sets
respectively over the baseline if the dropout rate of $r = 0.1$ is employed. 
From Tables \ref{tbl:sem_lower_bound_fixed}, \ref{tbl:sem_upper_bound_fixed}
and \ref{tbl:input_dropout} results, we conclude that the SEM algorithm
is much more effective than the conventional input dropout approach.
\begin{table}[!htbp]
  \renewcommand{\arraystretch}{1.3}
  \centering
        \caption{\label{tbl:sem_result_with_trans_lm}
        Word Error Rates (WERs) obtained with SEM processing
        using a modified shallow-fusion with a Transformer LM with different LM weights ($\lambda_{\text{lm}}$)
        }   
        \begin{tabular}{| c || c | c | c | c |}
          \hline
              \specialcell{ $\lambda_p$ \\ $ \lambda_{\text{lm}}$ } &   
              \specialcell{ 0.003\\ 0.36}  &
              \specialcell{ 0.003\\ 0.4}  &   
              \specialcell{ 0.003\\ 0.44}  &   
              \specialcell{ 0.003\\ 0.48} \\
              \hline
              test-clean  &   \textcolor{blue}{\textbf{2.52}}  \% & 2.62 \%  &   2.62 \% &  2.66 \% \\  
              test-other  &   7.93  \% & \textcolor{blue}{\textbf{7.87}} \%  &  \textcolor{blue}{\textbf{ 7.87 }} \% &  8.33 \% \\
              average     &   5.23  \% & \textcolor{blue}{\textbf{5.25}} \%  &  \textcolor{blue}{\textbf{  5.25 }} \% &  5.50 \% \\
          \hline
       \end{tabular}
       \vspace{-2mm}
\end{table}

 Using the modified shallow-fusion  introduced in \cite{c_kim_interspeech_2019_00} with a Transformer LM \cite{a_vaswani_nips_2017_00}, we obtain further improvement as shown in Table \ref{tbl:sem_result_with_trans_lm}.
$\lambda_p$ and $\lambda_{\text{lm}}$ in  Table
  \ref{tbl:sem_result_with_trans_lm} are weights for the prior probability and
  the LM prediction probability, respectively as in \cite{c_kim_interspeech_2019_00}.
When the beam size is 36, $\lambda_p$ is 0.003, and $\lambda_{\text{lm}}$ is
  $0.4 \sim 0.44$, we obtained a 2.62 \% WER on the {\tt test-clean} set and a 7.87 \% WER on the {\tt test-other} set.
\section{Conclusions}
  In this paper, we present the Small Energy Masking (SEM) algorithm and
  experimental results obtained using this algorithm.
In this algorithm, masks are created by comparing the energy in the
time-frequency bin and the energy threshold.
 A uniform distribution is employed to generate the ratio of this 
 energy threshold to the peak filterbank energy in each utterance in decibels.
  Unmasked feature elements are scaled so that the total sum of
  the feature values remain the same throughout this masking procedure.
Experimental results show that this algorithm shows relatively 11.2 \%
and 13.5 \% WER improvements on the standard
LibriSpeech test-clean and test-other sets over the baseline end-to-end
speech recognition system. Additionally, compared to the conventional input dropout 
algorithm, the SEM algorithm shows 7.7 \% and 11.6 \% relative
  improvements on the same LibriSpeech {\tt test-clean} and {\tt test-other} sets,
respectively.
  With a modified shallow-fusion technique with a Transformer LM
  \cite{c_kim_interspeech_2019_00},
we obtained a 2.62 \% WER on the LibriSpeech {\tt test-clean} set and a 7.87 \% WER
on the LibriSpeech {\tt test-other} set.
%
%
%
\bibliographystyle{IEEEtran}
\bibliography{common_bib_file}

\end{document}

%% file: sem_procedure_diagram.tex
\ifx\du\undefined
  \newlength{\du}
\fi
\setlength{\du}{15\unitlength}
\begin{tikzpicture}[even odd rule]
  \tikzstyle{every node}=[font={\Large\it}]
\pgftransformxscale{1.000000}
\pgftransformyscale{-1.000000}
\definecolor{dialinecolor}{rgb}{0.000000, 0.000000, 0.000000}
\pgfsetstrokecolor{dialinecolor}
\pgfsetstrokeopacity{1.000000}
\definecolor{diafillcolor}{rgb}{1.000000, 1.000000, 1.000000}
\pgfsetfillcolor{diafillcolor}
\pgfsetfillopacity{1.000000}
\pgfsetlinewidth{0.100000\du}
\pgfsetdash{{\pgflinewidth}{0.200000\du}}{0cm}
\pgfsetmiterjoin
{\pgfsetcornersarced{\pgfpoint{0.000000\du}{0.000000\du}}\definecolor{diafillcolor}{rgb}{1.000000, 1.000000, 1.000000}
\pgfsetfillcolor{diafillcolor}
\pgfsetfillopacity{1.000000}
\fill (32.000000\du,-2.000000\du)--(32.000000\du,22.000000\du)--(54.000000\du,22.000000\du)--(54.000000\du,-2.000000\du)--cycle;
}{\pgfsetcornersarced{\pgfpoint{0.000000\du}{0.000000\du}}\definecolor{dialinecolor}{rgb}{0.000000, 0.000000, 0.000000}
\pgfsetstrokecolor{dialinecolor}
\pgfsetstrokeopacity{1.000000}
\draw (32.000000\du,-2.000000\du)--(32.000000\du,22.000000\du)--(54.000000\du,22.000000\du)--(54.000000\du,-2.000000\du)--cycle;
}
\definecolor{dialinecolor}{rgb}{0.000000, 0.000000, 0.000000}
\pgfsetstrokecolor{dialinecolor}
\pgfsetstrokeopacity{1.000000}
\definecolor{diafillcolor}{rgb}{0.000000, 0.000000, 0.000000}
\pgfsetfillcolor{diafillcolor}
\pgfsetfillopacity{1.000000}
\node[anchor=base,inner sep=0pt, outer sep=0pt,color=dialinecolor] at (43.500000\du,10.195000\du){};
\pgfsetlinewidth{0.100000\du}
\pgfsetdash{}{0pt}
\pgfsetbuttcap
{
\definecolor{diafillcolor}{rgb}{0.000000, 0.000000, 0.000000}
\pgfsetfillcolor{diafillcolor}
\pgfsetfillopacity{1.000000}
\pgfsetarrowsend{stealth}
\definecolor{dialinecolor}{rgb}{0.000000, 0.000000, 0.000000}
\pgfsetstrokecolor{dialinecolor}
\pgfsetstrokeopacity{1.000000}
\draw (49.000000\du,24.000000\du)--(49.000000\du,21.000000\du);
}
\pgfsetlinewidth{0.100000\du}
\pgfsetdash{}{0pt}
\pgfsetbuttcap
{
\definecolor{diafillcolor}{rgb}{0.000000, 0.000000, 0.000000}
\pgfsetfillcolor{diafillcolor}
\pgfsetfillopacity{1.000000}
\pgfsetarrowsend{stealth}
\definecolor{dialinecolor}{rgb}{0.000000, 0.000000, 0.000000}
\pgfsetstrokecolor{dialinecolor}
\pgfsetstrokeopacity{1.000000}
\draw (49.000000\du,17.000000\du)--(49.000000\du,15.000000\du);
}
\pgfsetlinewidth{0.100000\du}
\pgfsetdash{}{0pt}
\pgfsetmiterjoin
{\pgfsetcornersarced{\pgfpoint{0.000000\du}{0.000000\du}}\definecolor{diafillcolor}{rgb}{0.000000, 1.000000, 0.000000}
\pgfsetfillcolor{diafillcolor}
\pgfsetfillopacity{1.000000}
\fill (45.000000\du,17.000000\du)--(45.000000\du,21.000000\du)--(53.000000\du,21.000000\du)--(53.000000\du,17.000000\du)--cycle;
}{\pgfsetcornersarced{\pgfpoint{0.000000\du}{0.000000\du}}\definecolor{dialinecolor}{rgb}{0.000000, 0.000000, 0.000000}
\pgfsetstrokecolor{dialinecolor}
\pgfsetstrokeopacity{1.000000}
\draw (45.000000\du,17.000000\du)--(45.000000\du,21.000000\du)--(53.000000\du,21.000000\du)--(53.000000\du,17.000000\du)--cycle;
}
\definecolor{dialinecolor}{rgb}{0.000000, 0.000000, 0.000000}
\pgfsetstrokecolor{dialinecolor}
\pgfsetstrokeopacity{1.000000}
\definecolor{diafillcolor}{rgb}{0.000000, 0.000000, 0.000000}
\pgfsetfillcolor{diafillcolor}
\pgfsetfillopacity{1.000000}
\node[anchor=base,inner sep=0pt, outer sep=0pt,color=dialinecolor] at (49.000000\du,18.795000\du){Peak Filterbank };
\definecolor{dialinecolor}{rgb}{0.000000, 0.000000, 0.000000}
\pgfsetstrokecolor{dialinecolor}
\pgfsetstrokeopacity{1.000000}
\definecolor{diafillcolor}{rgb}{0.000000, 0.000000, 0.000000}
\pgfsetfillcolor{diafillcolor}
\pgfsetfillopacity{1.000000}
\node[anchor=base,inner sep=0pt, outer sep=0pt,color=dialinecolor] at (49.000000\du,19.595000\du){Energy Calculation};
\pgfsetlinewidth{0.100000\du}
\pgfsetdash{}{0pt}
\pgfsetmiterjoin
{\pgfsetcornersarced{\pgfpoint{0.000000\du}{0.000000\du}}\definecolor{diafillcolor}{rgb}{0.117647, 0.564706, 1.000000}
\pgfsetfillcolor{diafillcolor}
\pgfsetfillopacity{1.000000}
\fill (56.000000\du,25.000000\du)--(56.000000\du,29.000000\du)--(64.000000\du,29.000000\du)--(64.000000\du,25.000000\du)--cycle;
}{\pgfsetcornersarced{\pgfpoint{0.000000\du}{0.000000\du}}\definecolor{dialinecolor}{rgb}{0.000000, 0.000000, 0.000000}
\pgfsetstrokecolor{dialinecolor}
\pgfsetstrokeopacity{1.000000}
\draw (56.000000\du,25.000000\du)--(56.000000\du,29.000000\du)--(64.000000\du,29.000000\du)--(64.000000\du,25.000000\du)--cycle;
}
\definecolor{dialinecolor}{rgb}{0.000000, 0.000000, 0.000000}
\pgfsetstrokecolor{dialinecolor}
\pgfsetstrokeopacity{1.000000}
\definecolor{diafillcolor}{rgb}{0.000000, 0.000000, 0.000000}
\pgfsetfillcolor{diafillcolor}
\pgfsetfillopacity{1.000000}
\node[anchor=base,inner sep=0pt, outer sep=0pt,color=dialinecolor] at (60.000000\du,26.795000\du){Filterbank };
\definecolor{dialinecolor}{rgb}{0.000000, 0.000000, 0.000000}
\pgfsetstrokecolor{dialinecolor}
\pgfsetstrokeopacity{1.000000}
\definecolor{diafillcolor}{rgb}{0.000000, 0.000000, 0.000000}
\pgfsetfillcolor{diafillcolor}
\pgfsetfillopacity{1.000000}
\node[anchor=base,inner sep=0pt, outer sep=0pt,color=dialinecolor] at (60.000000\du,27.595000\du){Energy Calculation};
\pgfsetlinewidth{0.100000\du}
\pgfsetdash{}{0pt}
\pgfsetbuttcap
{
\definecolor{diafillcolor}{rgb}{0.000000, 0.000000, 0.000000}
\pgfsetfillcolor{diafillcolor}
\pgfsetfillopacity{1.000000}
\pgfsetarrowsend{stealth}
\definecolor{dialinecolor}{rgb}{0.000000, 0.000000, 0.000000}
\pgfsetstrokecolor{dialinecolor}
\pgfsetstrokeopacity{1.000000}
\draw (60.000000\du,31.000000\du)--(60.000000\du,29.000000\du);
}
\pgfsetlinewidth{0.100000\du}
\pgfsetdash{}{0pt}
\pgfsetmiterjoin
{\pgfsetcornersarced{\pgfpoint{0.000000\du}{0.000000\du}}\definecolor{diafillcolor}{rgb}{0.000000, 1.000000, 0.000000}
\pgfsetfillcolor{diafillcolor}
\pgfsetfillopacity{1.000000}
\fill (45.000000\du,11.000000\du)--(45.000000\du,15.000000\du)--(53.000000\du,15.000000\du)--(53.000000\du,11.000000\du)--cycle;
}{\pgfsetcornersarced{\pgfpoint{0.000000\du}{0.000000\du}}\definecolor{dialinecolor}{rgb}{0.000000, 0.000000, 0.000000}
\pgfsetstrokecolor{dialinecolor}
\pgfsetstrokeopacity{1.000000}
\draw (45.000000\du,11.000000\du)--(45.000000\du,15.000000\du)--(53.000000\du,15.000000\du)--(53.000000\du,11.000000\du)--cycle;
}
\definecolor{dialinecolor}{rgb}{0.000000, 0.000000, 0.000000}
\pgfsetstrokecolor{dialinecolor}
\pgfsetstrokeopacity{1.000000}
\definecolor{diafillcolor}{rgb}{0.000000, 0.000000, 0.000000}
\pgfsetfillcolor{diafillcolor}
\pgfsetfillopacity{1.000000}
\node[anchor=base,inner sep=0pt, outer sep=0pt,color=dialinecolor] at (49.000000\du,12.395000\du){Filterbank };
\definecolor{dialinecolor}{rgb}{0.000000, 0.000000, 0.000000}
\pgfsetstrokecolor{dialinecolor}
\pgfsetstrokeopacity{1.000000}
\definecolor{diafillcolor}{rgb}{0.000000, 0.000000, 0.000000}
\pgfsetfillcolor{diafillcolor}
\pgfsetfillopacity{1.000000}
\node[anchor=base,inner sep=0pt, outer sep=0pt,color=dialinecolor] at (49.000000\du,13.195000\du){EnergyThreshold};
\definecolor{dialinecolor}{rgb}{0.000000, 0.000000, 0.000000}
\pgfsetstrokecolor{dialinecolor}
\pgfsetstrokeopacity{1.000000}
\definecolor{diafillcolor}{rgb}{0.000000, 0.000000, 0.000000}
\pgfsetfillcolor{diafillcolor}
\pgfsetfillopacity{1.000000}
\node[anchor=base,inner sep=0pt, outer sep=0pt,color=dialinecolor] at (49.000000\du,13.995000\du){Calculation};
\pgfsetlinewidth{0.100000\du}
\pgfsetdash{}{0pt}
\pgfsetbuttcap
{
\definecolor{diafillcolor}{rgb}{0.000000, 0.000000, 0.000000}
\pgfsetfillcolor{diafillcolor}
\pgfsetfillopacity{1.000000}
\pgfsetarrowsend{stealth}
\definecolor{dialinecolor}{rgb}{0.000000, 0.000000, 0.000000}
\pgfsetstrokecolor{dialinecolor}
\pgfsetstrokeopacity{1.000000}
\draw (41.522500\du,13.000000\du)--(45.000000\du,13.000000\du);
}
\pgfsetlinewidth{0.100000\du}
\pgfsetdash{}{0pt}
\pgfsetmiterjoin
{\pgfsetcornersarced{\pgfpoint{0.000000\du}{0.000000\du}}\definecolor{diafillcolor}{rgb}{0.000000, 1.000000, 0.000000}
\pgfsetfillcolor{diafillcolor}
\pgfsetfillopacity{1.000000}
\fill (33.500000\du,11.000000\du)--(33.500000\du,15.000000\du)--(41.522500\du,15.000000\du)--(41.522500\du,11.000000\du)--cycle;
}{\pgfsetcornersarced{\pgfpoint{0.000000\du}{0.000000\du}}\definecolor{dialinecolor}{rgb}{0.000000, 0.000000, 0.000000}
\pgfsetstrokecolor{dialinecolor}
\pgfsetstrokeopacity{1.000000}
\draw (33.500000\du,11.000000\du)--(33.500000\du,15.000000\du)--(41.522500\du,15.000000\du)--(41.522500\du,11.000000\du)--cycle;
}
\definecolor{dialinecolor}{rgb}{0.000000, 0.000000, 0.000000}
\pgfsetstrokecolor{dialinecolor}
\pgfsetstrokeopacity{1.000000}
\definecolor{diafillcolor}{rgb}{0.000000, 0.000000, 0.000000}
\pgfsetfillcolor{diafillcolor}
\pgfsetfillopacity{1.000000}
\node[anchor=base,inner sep=0pt, outer sep=0pt,color=dialinecolor] at
(37.5\du,12.795000\du){Uniform Random};
\definecolor{dialinecolor}{rgb}{0.000000, 0.000000, 0.000000}
\pgfsetstrokecolor{dialinecolor}
\pgfsetstrokeopacity{1.000000}
\definecolor{diafillcolor}{rgb}{0.000000, 0.000000, 0.000000}
\pgfsetfillcolor{diafillcolor}
\pgfsetfillopacity{1.000000}
\node[anchor=base,inner sep=0pt, outer sep=0pt,color=dialinecolor] at
(37.761250\du,13.595000\du){Generation};
\pgfsetlinewidth{0.100000\du}
\pgfsetdash{}{0pt}
\pgfsetbuttcap
{
\definecolor{diafillcolor}{rgb}{0.000000, 0.000000, 0.000000}
\pgfsetfillcolor{diafillcolor}
\pgfsetfillopacity{1.000000}
\pgfsetarrowsend{stealth}
\definecolor{dialinecolor}{rgb}{0.000000, 0.000000, 0.000000}
\pgfsetstrokecolor{dialinecolor}
\pgfsetstrokeopacity{1.000000}
\draw (49.000000\du,11.000000\du)--(49.000000\du,9.000000\du);
}
\pgfsetlinewidth{0.100000\du}
\pgfsetdash{}{0pt}
\pgfsetbuttcap
{
\definecolor{diafillcolor}{rgb}{0.000000, 0.000000, 0.000000}
\pgfsetfillcolor{diafillcolor}
\pgfsetfillopacity{1.000000}
\definecolor{dialinecolor}{rgb}{0.000000, 0.000000, 0.000000}
\pgfsetstrokecolor{dialinecolor}
\pgfsetstrokeopacity{1.000000}
\draw (49.000000\du,24.000000\du)--(60.000000\du,24.000000\du);
}
\pgfsetlinewidth{0.100000\du}
\pgfsetdash{}{0pt}
\pgfsetmiterjoin
{\pgfsetcornersarced{\pgfpoint{0.000000\du}{0.000000\du}}\definecolor{diafillcolor}{rgb}{0.000000, 1.000000, 0.000000}
\pgfsetfillcolor{diafillcolor}
\pgfsetfillopacity{1.000000}
\fill (45.000000\du,5.000000\du)--(45.000000\du,9.000000\du)--(53.000000\du,9.000000\du)--(53.000000\du,5.000000\du)--cycle;
}{\pgfsetcornersarced{\pgfpoint{0.000000\du}{0.000000\du}}\definecolor{dialinecolor}{rgb}{0.000000, 0.000000, 0.000000}
\pgfsetstrokecolor{dialinecolor}
\pgfsetstrokeopacity{1.000000}
\draw (45.000000\du,5.000000\du)--(45.000000\du,9.000000\du)--(53.000000\du,9.000000\du)--(53.000000\du,5.000000\du)--cycle;
}
\definecolor{dialinecolor}{rgb}{0.000000, 0.000000, 0.000000}
\pgfsetstrokecolor{dialinecolor}
\pgfsetstrokeopacity{1.000000}
\definecolor{diafillcolor}{rgb}{0.000000, 0.000000, 0.000000}
\pgfsetfillcolor{diafillcolor}
\pgfsetfillopacity{1.000000}
\node[anchor=base,inner sep=0pt, outer sep=0pt,color=dialinecolor] at (49.000000\du,6.795000\du){Binary Mask};
\definecolor{dialinecolor}{rgb}{0.000000, 0.000000, 0.000000}
\pgfsetstrokecolor{dialinecolor}
\pgfsetstrokeopacity{1.000000}
\definecolor{diafillcolor}{rgb}{0.000000, 0.000000, 0.000000}
\pgfsetfillcolor{diafillcolor}
\pgfsetfillopacity{1.000000}
\node[anchor=base,inner sep=0pt, outer sep=0pt,color=dialinecolor] at (49.000000\du,7.595000\du){Construction};
\pgfsetlinewidth{0.100000\du}
\pgfsetdash{}{0pt}
\pgfsetbuttcap
{
\definecolor{diafillcolor}{rgb}{0.000000, 0.000000, 0.000000}
\pgfsetfillcolor{diafillcolor}
\pgfsetfillopacity{1.000000}
\pgfsetarrowsend{stealth}
\definecolor{dialinecolor}{rgb}{0.000000, 0.000000, 0.000000}
\pgfsetstrokecolor{dialinecolor}
\pgfsetstrokeopacity{1.000000}
\draw (60.000000\du,24.000000\du)--(60.000000\du,21.000000\du);
}
\pgfsetlinewidth{0.100000\du}
\pgfsetdash{}{0pt}
\pgfsetmiterjoin
{\pgfsetcornersarced{\pgfpoint{0.000000\du}{0.000000\du}}\definecolor{diafillcolor}{rgb}{0.117647, 0.564706, 1.000000}
\pgfsetfillcolor{diafillcolor}
\pgfsetfillopacity{1.000000}
\fill (56.000000\du,17.000000\du)--(56.000000\du,21.000000\du)--(64.000000\du,21.000000\du)--(64.000000\du,17.000000\du)--cycle;
}{\pgfsetcornersarced{\pgfpoint{0.000000\du}{0.000000\du}}\definecolor{dialinecolor}{rgb}{0.000000, 0.000000, 0.000000}
\pgfsetstrokecolor{dialinecolor}
\pgfsetstrokeopacity{1.000000}
\draw (56.000000\du,17.000000\du)--(56.000000\du,21.000000\du)--(64.000000\du,21.000000\du)--(64.000000\du,17.000000\du)--cycle;
}
\definecolor{dialinecolor}{rgb}{0.000000, 0.000000, 0.000000}
\pgfsetstrokecolor{dialinecolor}
\pgfsetstrokeopacity{1.000000}
\definecolor{diafillcolor}{rgb}{0.000000, 0.000000, 0.000000}
\pgfsetfillcolor{diafillcolor}
\pgfsetfillopacity{1.000000}
\node[anchor=base,inner sep=0pt, outer sep=0pt,color=dialinecolor] at (60.000000\du,18.795000\du){Power-Law};
\definecolor{dialinecolor}{rgb}{0.000000, 0.000000, 0.000000}
\pgfsetstrokecolor{dialinecolor}
\pgfsetstrokeopacity{1.000000}
\definecolor{diafillcolor}{rgb}{0.000000, 0.000000, 0.000000}
\pgfsetfillcolor{diafillcolor}
\pgfsetfillopacity{1.000000}
\node[anchor=base,inner sep=0pt, outer sep=0pt,color=dialinecolor] at (60.000000\du,19.595000\du){Nonlinearity};
\pgfsetlinewidth{0.100000\du}
\pgfsetdash{}{0pt}
\pgfsetbuttcap
{
\definecolor{diafillcolor}{rgb}{0.000000, 0.000000, 0.000000}
\pgfsetfillcolor{diafillcolor}
\pgfsetfillopacity{1.000000}
\pgfsetarrowsend{stealth}
\definecolor{dialinecolor}{rgb}{0.000000, 0.000000, 0.000000}
\pgfsetstrokecolor{dialinecolor}
\pgfsetstrokeopacity{1.000000}
\draw (60.000000\du,17.000000\du)--(60.000000\du,15.000000\du);
}
\pgfsetlinewidth{0.100000\du}
\pgfsetdash{}{0pt}
\pgfsetbuttcap
{
\definecolor{diafillcolor}{rgb}{0.000000, 0.000000, 0.000000}
\pgfsetfillcolor{diafillcolor}
\pgfsetfillopacity{1.000000}
\pgfsetarrowsend{stealth}
\definecolor{dialinecolor}{rgb}{0.000000, 0.000000, 0.000000}
\pgfsetstrokecolor{dialinecolor}
\pgfsetstrokeopacity{1.000000}
\draw (53.000000\du,7.000000\du)--(59.000000\du,7.000000\du);
}
\pgfsetlinewidth{0.100000\du}
\pgfsetdash{}{0pt}
\pgfsetbuttcap
{
\definecolor{diafillcolor}{rgb}{0.000000, 0.000000, 0.000000}
\pgfsetfillcolor{diafillcolor}
\pgfsetfillopacity{1.000000}
\pgfsetarrowsend{stealth}
\definecolor{dialinecolor}{rgb}{0.000000, 0.000000, 0.000000}
\pgfsetstrokecolor{dialinecolor}
\pgfsetstrokeopacity{1.000000}
\draw (60.000000\du,5.949890\du)--(60.000000\du,2.000000\du);
}
\pgfsetlinewidth{0.100000\du}
\pgfsetdash{}{0pt}
\pgfsetbuttcap
{
\definecolor{diafillcolor}{rgb}{0.000000, 0.000000, 0.000000}
\pgfsetfillcolor{diafillcolor}
\pgfsetfillopacity{1.000000}
\definecolor{dialinecolor}{rgb}{0.000000, 0.000000, 0.000000}
\pgfsetstrokecolor{dialinecolor}
\pgfsetstrokeopacity{1.000000}
\draw (60.000000\du,25.000000\du)--(60.000000\du,24.000000\du);
}
\pgfsetlinewidth{0.100000\du}
\pgfsetdash{}{0pt}
\pgfsetmiterjoin
{\pgfsetcornersarced{\pgfpoint{0.000000\du}{0.000000\du}}\definecolor{diafillcolor}{rgb}{0.117647, 0.564706, 1.000000}
\pgfsetfillcolor{diafillcolor}
\pgfsetfillopacity{1.000000}
\fill (56.000000\du,11.000000\du)--(56.000000\du,15.000000\du)--(64.000000\du,15.000000\du)--(64.000000\du,11.000000\du)--cycle;
}{\pgfsetcornersarced{\pgfpoint{0.000000\du}{0.000000\du}}\definecolor{dialinecolor}{rgb}{0.000000, 0.000000, 0.000000}
\pgfsetstrokecolor{dialinecolor}
\pgfsetstrokeopacity{1.000000}
\draw (56.000000\du,11.000000\du)--(56.000000\du,15.000000\du)--(64.000000\du,15.000000\du)--(64.000000\du,11.000000\du)--cycle;
}
\definecolor{dialinecolor}{rgb}{0.000000, 0.000000, 0.000000}
\pgfsetstrokecolor{dialinecolor}
\pgfsetstrokeopacity{1.000000}
\definecolor{diafillcolor}{rgb}{0.000000, 0.000000, 0.000000}
\pgfsetfillcolor{diafillcolor}
\pgfsetfillopacity{1.000000}
\node[anchor=base,inner sep=0pt, outer sep=0pt,color=dialinecolor] at (60.000000\du,12.395000\du){Global Mean and};
\definecolor{dialinecolor}{rgb}{0.000000, 0.000000, 0.000000}
\pgfsetstrokecolor{dialinecolor}
\pgfsetstrokeopacity{1.000000}
\definecolor{diafillcolor}{rgb}{0.000000, 0.000000, 0.000000}
\pgfsetfillcolor{diafillcolor}
\pgfsetfillopacity{1.000000}
\node[anchor=base,inner sep=0pt, outer sep=0pt,color=dialinecolor] at (60.000000\du,13.195000\du){Variannce };
\definecolor{dialinecolor}{rgb}{0.000000, 0.000000, 0.000000}
\pgfsetstrokecolor{dialinecolor}
\pgfsetstrokeopacity{1.000000}
\definecolor{diafillcolor}{rgb}{0.000000, 0.000000, 0.000000}
\pgfsetfillcolor{diafillcolor}
\pgfsetfillopacity{1.000000}
\node[anchor=base,inner sep=0pt, outer sep=0pt,color=dialinecolor] at (60.000000\du,13.995000\du){Normalization};
\pgfsetlinewidth{0.100000\du}
\pgfsetdash{}{0pt}
\pgfsetbuttcap
{
\definecolor{diafillcolor}{rgb}{0.000000, 0.000000, 0.000000}
\pgfsetfillcolor{diafillcolor}
\pgfsetfillopacity{1.000000}
\pgfsetarrowsend{stealth}
\definecolor{dialinecolor}{rgb}{0.000000, 0.000000, 0.000000}
\pgfsetstrokecolor{dialinecolor}
\pgfsetstrokeopacity{1.000000}
\draw (60.000000\du,11.000000\du)--(60.000000\du,8.000000\du);
}
\pgfsetlinewidth{0.100000\du}
\pgfsetdash{}{0pt}
\pgfsetbuttcap
{
\definecolor{diafillcolor}{rgb}{0.000000, 0.000000, 0.000000}
\pgfsetfillcolor{diafillcolor}
\pgfsetfillopacity{1.000000}
\pgfsetarrowsend{stealth}
\definecolor{dialinecolor}{rgb}{0.000000, 0.000000, 0.000000}
\pgfsetstrokecolor{dialinecolor}
\pgfsetstrokeopacity{1.000000}
\draw (60.000000\du,0.000000\du)--(60.000000\du,-3.000000\du);
}
\pgfsetlinewidth{0.100000\du}
\pgfsetdash{}{0pt}
\pgfsetmiterjoin
\definecolor{diafillcolor}{rgb}{1.000000, 1.000000, 1.000000}
\pgfsetfillcolor{diafillcolor}
\pgfsetfillopacity{1.000000}
\pgfpathellipse{\pgfpoint{60.000000\du}{7.000000\du}}{\pgfpoint{1.000000\du}{0\du}}{\pgfpoint{0\du}{1.000000\du}}
\pgfusepath{fill}
\definecolor{dialinecolor}{rgb}{0.000000, 0.000000, 0.000000}
\pgfsetstrokecolor{dialinecolor}
\pgfsetstrokeopacity{1.000000}
\pgfpathellipse{\pgfpoint{60.000000\du}{7.000000\du}}{\pgfpoint{1.000000\du}{0\du}}{\pgfpoint{0\du}{1.000000\du}}
\pgfusepath{stroke}
\definecolor{dialinecolor}{rgb}{0.000000, 0.000000, 0.000000}
\pgfsetstrokecolor{dialinecolor}
\pgfsetstrokeopacity{1.000000}
\definecolor{diafillcolor}{rgb}{0.000000, 0.000000, 0.000000}
\pgfsetfillcolor{diafillcolor}
\pgfsetfillopacity{1.000000}
\node[anchor=base,inner sep=0pt, outer sep=0pt,color=dialinecolor] at (60.000000\du,7.195000\du){};
\pgfsetlinewidth{0.100000\du}
\pgfsetdash{}{0pt}
\pgfsetbuttcap
{
\definecolor{diafillcolor}{rgb}{0.000000, 0.000000, 0.000000}
\pgfsetfillcolor{diafillcolor}
\pgfsetfillopacity{1.000000}
\definecolor{dialinecolor}{rgb}{0.000000, 0.000000, 0.000000}
\pgfsetstrokecolor{dialinecolor}
\pgfsetstrokeopacity{1.000000}
\draw (59.292900\du,7.707110\du)--(60.707100\du,6.292890\du);
}
\pgfsetlinewidth{0.100000\du}
\pgfsetdash{}{0pt}
\pgfsetbuttcap
{
\definecolor{diafillcolor}{rgb}{0.000000, 0.000000, 0.000000}
\pgfsetfillcolor{diafillcolor}
\pgfsetfillopacity{1.000000}
\definecolor{dialinecolor}{rgb}{0.000000, 0.000000, 0.000000}
\pgfsetstrokecolor{dialinecolor}
\pgfsetstrokeopacity{1.000000}
\draw (60.707100\du,7.707110\du)--(59.292900\du,6.292890\du);
}
\pgfsetlinewidth{0.100000\du}
\pgfsetdash{}{0pt}
\pgfsetmiterjoin
\definecolor{diafillcolor}{rgb}{1.000000, 1.000000, 1.000000}
\pgfsetfillcolor{diafillcolor}
\pgfsetfillopacity{1.000000}
\pgfpathellipse{\pgfpoint{60.000000\du}{1.000000\du}}{\pgfpoint{1.000000\du}{0\du}}{\pgfpoint{0\du}{1.000000\du}}
\pgfusepath{fill}
\definecolor{dialinecolor}{rgb}{0.000000, 0.000000, 0.000000}
\pgfsetstrokecolor{dialinecolor}
\pgfsetstrokeopacity{1.000000}
\pgfpathellipse{\pgfpoint{60.000000\du}{1.000000\du}}{\pgfpoint{1.000000\du}{0\du}}{\pgfpoint{0\du}{1.000000\du}}
\pgfusepath{stroke}
\definecolor{dialinecolor}{rgb}{0.000000, 0.000000, 0.000000}
\pgfsetstrokecolor{dialinecolor}
\pgfsetstrokeopacity{1.000000}
\definecolor{diafillcolor}{rgb}{0.000000, 0.000000, 0.000000}
\pgfsetfillcolor{diafillcolor}
\pgfsetfillopacity{1.000000}
\node[anchor=base,inner sep=0pt, outer sep=0pt,color=dialinecolor] at (60.000000\du,1.195000\du){};
\pgfsetlinewidth{0.100000\du}
\pgfsetdash{}{0pt}
\pgfsetbuttcap
{
\definecolor{diafillcolor}{rgb}{0.000000, 0.000000, 0.000000}
\pgfsetfillcolor{diafillcolor}
\pgfsetfillopacity{1.000000}
\definecolor{dialinecolor}{rgb}{0.000000, 0.000000, 0.000000}
\pgfsetstrokecolor{dialinecolor}
\pgfsetstrokeopacity{1.000000}
\draw (59.292900\du,1.707110\du)--(60.707100\du,0.292893\du);
}
\pgfsetlinewidth{0.100000\du}
\pgfsetdash{}{0pt}
\pgfsetbuttcap
{
\definecolor{diafillcolor}{rgb}{0.000000, 0.000000, 0.000000}
\pgfsetfillcolor{diafillcolor}
\pgfsetfillopacity{1.000000}
\definecolor{dialinecolor}{rgb}{0.000000, 0.000000, 0.000000}
\pgfsetstrokecolor{dialinecolor}
\pgfsetstrokeopacity{1.000000}
\draw (60.707100\du,1.707110\du)--(59.292900\du,0.292893\du);
}
\pgfsetlinewidth{0.100000\du}
\pgfsetdash{}{0pt}
\pgfsetbuttcap
{
\definecolor{diafillcolor}{rgb}{0.000000, 0.000000, 0.000000}
\pgfsetfillcolor{diafillcolor}
\pgfsetfillopacity{1.000000}
\pgfsetarrowsend{stealth}
\definecolor{dialinecolor}{rgb}{0.000000, 0.000000, 0.000000}
\pgfsetstrokecolor{dialinecolor}
\pgfsetstrokeopacity{1.000000}
\draw (53.000000\du,1.000000\du)--(59.000000\du,1.000000\du);
}
\pgfsetlinewidth{0.100000\du}
\pgfsetdash{}{0pt}
\pgfsetmiterjoin
{\pgfsetcornersarced{\pgfpoint{0.000000\du}{0.000000\du}}\definecolor{diafillcolor}{rgb}{0.000000, 1.000000, 0.000000}
\pgfsetfillcolor{diafillcolor}
\pgfsetfillopacity{1.000000}
\fill (45.000000\du,-1.150000\du)--(45.000000\du,3.150000\du)--(53.000000\du,3.150000\du)--(53.000000\du,-1.150000\du)--cycle;
}{\pgfsetcornersarced{\pgfpoint{0.000000\du}{0.000000\du}}\definecolor{dialinecolor}{rgb}{0.000000, 0.000000, 0.000000}
\pgfsetstrokecolor{dialinecolor}
\pgfsetstrokeopacity{1.000000}
\draw (45.000000\du,-1.150000\du)--(45.000000\du,3.150000\du)--(53.000000\du,3.150000\du)--(53.000000\du,-1.150000\du)--cycle;
}
\definecolor{dialinecolor}{rgb}{0.000000, 0.000000, 0.000000}
\pgfsetstrokecolor{dialinecolor}
\pgfsetstrokeopacity{1.000000}
\definecolor{diafillcolor}{rgb}{0.000000, 0.000000, 0.000000}
\pgfsetfillcolor{diafillcolor}
\pgfsetfillopacity{1.000000}
\node[anchor=base,inner sep=0pt, outer sep=0pt,color=dialinecolor] at (49.000000\du,0.795000\du){Scaling Up to };
\definecolor{dialinecolor}{rgb}{0.000000, 0.000000, 0.000000}
\pgfsetstrokecolor{dialinecolor}
\pgfsetstrokeopacity{1.000000}
\definecolor{diafillcolor}{rgb}{0.000000, 0.000000, 0.000000}
\pgfsetfillcolor{diafillcolor}
\pgfsetfillopacity{1.000000}
\node[anchor=base,inner sep=0pt, outer sep=0pt,color=dialinecolor] at (49.000000\du,1.595000\du){Keep the Sum};
\definecolor{dialinecolor}{rgb}{0.000000, 0.000000, 0.000000}
\pgfsetstrokecolor{dialinecolor}
\pgfsetstrokeopacity{1.000000}
\definecolor{diafillcolor}{rgb}{0.000000, 0.000000, 0.000000}
\pgfsetfillcolor{diafillcolor}
\pgfsetfillopacity{1.000000}
\node[anchor=base west,inner sep=0pt,outer sep=0pt,color=dialinecolor] at
(60.945475\du,10.000000\du){input feature $\vec{x}\,[m]$};
\definecolor{dialinecolor}{rgb}{0.000000, 0.000000, 0.000000}
\pgfsetstrokecolor{dialinecolor}
\pgfsetstrokeopacity{1.000000}
\definecolor{diafillcolor}{rgb}{0.000000, 0.000000, 0.000000}
\pgfsetfillcolor{diafillcolor}
\pgfsetfillopacity{1.000000}
\node[anchor=base west,inner sep=0pt,outer sep=0pt,color=dialinecolor] at (60.945475\du,30.272627\du){speech};
\definecolor{dialinecolor}{rgb}{0.000000, 0.000000, 0.000000}
\pgfsetstrokecolor{dialinecolor}
\pgfsetstrokeopacity{1.000000}
\definecolor{diafillcolor}{rgb}{0.000000, 0.000000, 0.000000}
\pgfsetfillcolor{diafillcolor}
\pgfsetfillopacity{1.000000}
\node[anchor=base west,inner sep=0pt,outer sep=0pt,color=dialinecolor] at (61.000000\du,-1.000000\du){masked feature};
\definecolor{dialinecolor}{rgb}{0.000000, 0.000000, 0.000000}
\pgfsetstrokecolor{dialinecolor}
\pgfsetstrokeopacity{1.000000}
\definecolor{diafillcolor}{rgb}{0.000000, 0.000000, 0.000000}
\pgfsetfillcolor{diafillcolor}
\pgfsetfillopacity{1.000000}
\node[anchor=base west,inner sep=0pt,outer sep=0pt,color=dialinecolor] at (55.000000\du,0.500000\du){r};
\definecolor{dialinecolor}{rgb}{0.000000, 0.000000, 0.000000}
\pgfsetstrokecolor{dialinecolor}
\pgfsetstrokeopacity{1.000000}
\definecolor{diafillcolor}{rgb}{0.000000, 0.000000, 0.000000}
\pgfsetfillcolor{diafillcolor}
\pgfsetfillopacity{1.000000}
\node[anchor=base west,inner sep=0pt,outer sep=0pt,color=dialinecolor] at
(60.500000\du,23.000000\du){$\vec{e}\,[m]$};
\definecolor{dialinecolor}{rgb}{0.000000, 0.000000, 0.000000}
\pgfsetstrokecolor{dialinecolor}
\pgfsetstrokeopacity{1.000000}
\definecolor{diafillcolor}{rgb}{0.000000, 0.000000, 0.000000}
\pgfsetfillcolor{diafillcolor}
\pgfsetfillopacity{1.000000}
\node[anchor=base west,inner sep=0pt,outer sep=0pt,color=dialinecolor] at
(55.000000\du,8.000000\du){$\mu[m, c]$};
\definecolor{dialinecolor}{rgb}{0.000000, 0.000000, 0.000000}
\pgfsetstrokecolor{dialinecolor}
\pgfsetstrokeopacity{1.000000}
\definecolor{diafillcolor}{rgb}{0.000000, 0.000000, 0.000000}
\pgfsetfillcolor{diafillcolor}
\pgfsetfillopacity{1.000000}
\node[anchor=base west,inner sep=0pt,outer sep=0pt,color=dialinecolor] at
(49.727373\du,16.272627\du){$e_{\text{peak}}$};
\definecolor{dialinecolor}{rgb}{0.000000, 0.000000, 0.000000}
\pgfsetstrokecolor{dialinecolor}
\pgfsetstrokeopacity{1.000000}
\definecolor{diafillcolor}{rgb}{0.000000, 0.000000, 0.000000}
\pgfsetfillcolor{diafillcolor}
\pgfsetfillopacity{1.000000}
\node[anchor=base west,inner sep=0pt,outer sep=0pt,color=dialinecolor] at
(49.500000\du,10.218102\du){$e_{th}$};
\definecolor{dialinecolor}{rgb}{0.000000, 0.000000, 0.000000}
\pgfsetstrokecolor{dialinecolor}
\pgfsetstrokeopacity{1.000000}
\definecolor{diafillcolor}{rgb}{0.000000, 0.000000, 0.000000}
\pgfsetfillcolor{diafillcolor}
\pgfsetfillopacity{1.000000}
\node[anchor=base west,inner sep=0pt,outer sep=0pt,color=dialinecolor] at
(42.000000\du,14.000000\du){$\eta_{th}$};
\definecolor{dialinecolor}{rgb}{0.000000, 0.000000, 0.000000}
\pgfsetstrokecolor{dialinecolor}
\pgfsetstrokeopacity{1.000000}
\definecolor{diafillcolor}{rgb}{0.000000, 0.000000, 0.000000}
\pgfsetfillcolor{diafillcolor}
\pgfsetfillopacity{1.000000}
\node[anchor=base west,inner sep=0pt,outer sep=0pt,color=dialinecolor] at
(60.500000\du,16.000000\du){$\vec{p}\,[m]$};
\definecolor{dialinecolor}{rgb}{0.000000, 0.000000, 0.000000}
\pgfsetstrokecolor{dialinecolor}
\pgfsetstrokeopacity{1.000000}
\definecolor{diafillcolor}{rgb}{0.000000, 0.000000, 0.000000}
\pgfsetfillcolor{diafillcolor}
\pgfsetfillopacity{1.000000}
\definecolor{dialinecolor}{rgb}{0.000000, 0.000000, 0.000000}
\pgfsetstrokecolor{dialinecolor}
\pgfsetstrokeopacity{1.000000}
\definecolor{diafillcolor}{rgb}{0.000000, 0.000000, 0.000000}
\pgfsetfillcolor{diafillcolor}
\pgfsetfillopacity{1.000000}
\node[anchor=base west,inner sep=0pt,outer sep=0pt,color=dialinecolor] at (61.000000\du,9.800000\du){};
\end{tikzpicture}

%% file: e2e_asr_block_diagram.tex
\ifx\du\undefined
  \newlength{\du}
\fi
\setlength{\du}{15\unitlength}
\begin{tikzpicture}[even odd rule]
  \tikzstyle{every node}=[font=\Large\itshape]
\pgftransformxscale{1.000000}
\pgftransformyscale{-1.000000}
\definecolor{dialinecolor}{rgb}{0.000000, 0.000000, 0.000000}
\pgfsetstrokecolor{dialinecolor}
\pgfsetstrokeopacity{1.000000}
\definecolor{diafillcolor}{rgb}{1.000000, 1.000000, 1.000000}
\pgfsetfillcolor{diafillcolor}
\pgfsetfillopacity{1.000000}
\pgfsetlinewidth{0.100000\du}
\pgfsetdash{}{0pt}
\pgfsetmiterjoin
\definecolor{diafillcolor}{rgb}{1.000000, 1.000000, 1.000000}
\pgfsetfillcolor{diafillcolor}
\pgfsetfillopacity{1.000000}
\pgfpathellipse{\pgfpoint{17.913943\du}{-6.956820\du}}{\pgfpoint{1.258243\du}{0\du}}{\pgfpoint{0\du}{1.250000\du}}
\pgfusepath{fill}
\definecolor{dialinecolor}{rgb}{0.000000, 0.000000, 0.000000}
\pgfsetstrokecolor{dialinecolor}
\pgfsetstrokeopacity{1.000000}
\pgfpathellipse{\pgfpoint{17.913943\du}{-6.956820\du}}{\pgfpoint{1.258243\du}{0\du}}{\pgfpoint{0\du}{1.250000\du}}
\pgfusepath{stroke}
\definecolor{dialinecolor}{rgb}{0.000000, 0.000000, 0.000000}
\pgfsetstrokecolor{dialinecolor}
\pgfsetstrokeopacity{1.000000}
\definecolor{diafillcolor}{rgb}{0.000000, 0.000000, 0.000000}
\pgfsetfillcolor{diafillcolor}
\pgfsetfillopacity{1.000000}
\node[anchor=base,inner sep=0pt, outer sep=0pt,color=dialinecolor] at (17.913943\du,-6.761820\du){};
\pgfsetlinewidth{0.100000\du}
\pgfsetdash{}{0pt}
\pgfsetmiterjoin
\definecolor{diafillcolor}{rgb}{1.000000, 1.000000, 1.000000}
\pgfsetfillcolor{diafillcolor}
\pgfsetfillopacity{1.000000}
\pgfpathellipse{\pgfpoint{25.871843\du}{5.950110\du}}{\pgfpoint{1.258243\du}{0\du}}{\pgfpoint{0\du}{1.250000\du}}
\pgfusepath{fill}
\definecolor{dialinecolor}{rgb}{0.000000, 0.000000, 0.000000}
\pgfsetstrokecolor{dialinecolor}
\pgfsetstrokeopacity{1.000000}
\pgfpathellipse{\pgfpoint{25.871843\du}{5.950110\du}}{\pgfpoint{1.258243\du}{0\du}}{\pgfpoint{0\du}{1.250000\du}}
\pgfusepath{stroke}
\definecolor{dialinecolor}{rgb}{0.000000, 0.000000, 0.000000}
\pgfsetstrokecolor{dialinecolor}
\pgfsetstrokeopacity{1.000000}
\definecolor{diafillcolor}{rgb}{0.000000, 0.000000, 0.000000}
\pgfsetfillcolor{diafillcolor}
\pgfsetfillopacity{1.000000}
\node[anchor=base,inner sep=0pt, outer sep=0pt,color=dialinecolor] at (25.871843\du,6.145110\du){};
\pgfsetlinewidth{0.100000\du}
\pgfsetdash{}{0pt}
\pgfsetmiterjoin
\definecolor{diafillcolor}{rgb}{1.000000, 1.000000, 1.000000}
\pgfsetfillcolor{diafillcolor}
\pgfsetfillopacity{1.000000}
\pgfpathellipse{\pgfpoint{25.849843\du}{14.960800\du}}{\pgfpoint{1.258243\du}{0\du}}{\pgfpoint{0\du}{1.250000\du}}
\pgfusepath{fill}
\definecolor{dialinecolor}{rgb}{0.000000, 0.000000, 0.000000}
\pgfsetstrokecolor{dialinecolor}
\pgfsetstrokeopacity{1.000000}
\pgfpathellipse{\pgfpoint{25.849843\du}{14.960800\du}}{\pgfpoint{1.258243\du}{0\du}}{\pgfpoint{0\du}{1.250000\du}}
\pgfusepath{stroke}
\definecolor{dialinecolor}{rgb}{0.000000, 0.000000, 0.000000}
\pgfsetstrokecolor{dialinecolor}
\pgfsetstrokeopacity{1.000000}
\definecolor{diafillcolor}{rgb}{0.000000, 0.000000, 0.000000}
\pgfsetfillcolor{diafillcolor}
\pgfsetfillopacity{1.000000}
\node[anchor=base,inner sep=0pt, outer sep=0pt,color=dialinecolor] at (25.849843\du,15.155800\du){};
\pgfsetlinewidth{0.100000\du}
\pgfsetdash{}{0pt}
\pgfsetmiterjoin
{\pgfsetcornersarced{\pgfpoint{0.000000\du}{0.000000\du}}\definecolor{diafillcolor}{rgb}{1.000000, 0.647059, 0.000000}
\pgfsetfillcolor{diafillcolor}
\pgfsetfillopacity{1.000000}
\fill (19.000000\du,17.000000\du)--(19.000000\du,19.050000\du)--(25.000000\du,19.050000\du)--(25.000000\du,17.000000\du)--cycle;
}{\pgfsetcornersarced{\pgfpoint{0.000000\du}{0.000000\du}}\definecolor{dialinecolor}{rgb}{0.000000, 0.000000, 0.000000}
\pgfsetstrokecolor{dialinecolor}
\pgfsetstrokeopacity{1.000000}
\draw (19.000000\du,17.000000\du)--(19.000000\du,19.050000\du)--(25.000000\du,19.050000\du)--(25.000000\du,17.000000\du)--cycle;
}
\definecolor{dialinecolor}{rgb}{0.000000, 0.000000, 0.000000}
\pgfsetstrokecolor{dialinecolor}
\pgfsetstrokeopacity{1.000000}
\definecolor{diafillcolor}{rgb}{0.000000, 0.000000, 0.000000}
\pgfsetfillcolor{diafillcolor}
\pgfsetfillopacity{1.000000}
\node[anchor=base,inner sep=0pt, outer sep=0pt,color=dialinecolor] at (22.000000\du,18.220000\du){MAXPOOL};
\pgfsetlinewidth{0.100000\du}
\pgfsetdash{}{0pt}
\pgfsetmiterjoin
{\pgfsetcornersarced{\pgfpoint{0.000000\du}{0.000000\du}}\definecolor{diafillcolor}{rgb}{0.000000, 1.000000, 0.000000}
\pgfsetfillcolor{diafillcolor}
\pgfsetfillopacity{1.000000}
\fill (19.000000\du,5.000000\du)--(19.000000\du,7.050000\du)--(25.000000\du,7.050000\du)--(25.000000\du,5.000000\du)--cycle;
}{\pgfsetcornersarced{\pgfpoint{0.000000\du}{0.000000\du}}\definecolor{dialinecolor}{rgb}{0.000000, 0.000000, 0.000000}
\pgfsetstrokecolor{dialinecolor}
\pgfsetstrokeopacity{1.000000}
\draw (19.000000\du,5.000000\du)--(19.000000\du,7.050000\du)--(25.000000\du,7.050000\du)--(25.000000\du,5.000000\du)--cycle;
}
\definecolor{dialinecolor}{rgb}{0.000000, 0.000000, 0.000000}
\pgfsetstrokecolor{dialinecolor}
\pgfsetstrokeopacity{1.000000}
\definecolor{diafillcolor}{rgb}{0.000000, 0.000000, 0.000000}
\pgfsetfillcolor{diafillcolor}
\pgfsetfillopacity{1.000000}
\node[anchor=base,inner sep=0pt, outer sep=0pt,color=dialinecolor] at (22.000000\du,6.220000\du){LSTM};
\pgfsetlinewidth{0.100000\du}
\pgfsetdash{}{0pt}
\pgfsetmiterjoin
{\pgfsetcornersarced{\pgfpoint{0.000000\du}{0.000000\du}}\definecolor{diafillcolor}{rgb}{1.000000, 0.000000, 0.000000}
\pgfsetfillcolor{diafillcolor}
\pgfsetfillopacity{1.000000}
\fill (16.000000\du,-1.000000\du)--(16.000000\du,1.050000\du)--(24.000000\du,1.050000\du)--(24.000000\du,-1.000000\du)--cycle;
}{\pgfsetcornersarced{\pgfpoint{0.000000\du}{0.000000\du}}\definecolor{dialinecolor}{rgb}{0.000000, 0.000000, 0.000000}
\pgfsetstrokecolor{dialinecolor}
\pgfsetstrokeopacity{1.000000}
\draw (16.000000\du,-1.000000\du)--(16.000000\du,1.050000\du)--(24.000000\du,1.050000\du)--(24.000000\du,-1.000000\du)--cycle;
}
\definecolor{dialinecolor}{rgb}{0.000000, 0.000000, 0.000000}
\pgfsetstrokecolor{dialinecolor}
\pgfsetstrokeopacity{1.000000}
\definecolor{diafillcolor}{rgb}{0.000000, 0.000000, 0.000000}
\pgfsetfillcolor{diafillcolor}
\pgfsetfillopacity{1.000000}
\node[anchor=base,inner sep=0pt, outer sep=0pt,color=dialinecolor] at (20.000000\du,0.220000\du){Attention};
\pgfsetlinewidth{0.100000\du}
\pgfsetdash{}{0pt}
\pgfsetmiterjoin
{\pgfsetcornersarced{\pgfpoint{0.000000\du}{0.000000\du}}\definecolor{diafillcolor}{rgb}{1.000000, 0.647059, 0.000000}
\pgfsetfillcolor{diafillcolor}
\pgfsetfillopacity{1.000000}
\fill (11.000000\du,-8.000000\du)--(11.000000\du,-6.000000\du)--(17.000000\du,-6.000000\du)--(17.000000\du,-8.000000\du)--cycle;
}{\pgfsetcornersarced{\pgfpoint{0.000000\du}{0.000000\du}}\definecolor{dialinecolor}{rgb}{0.000000, 0.000000, 0.000000}
\pgfsetstrokecolor{dialinecolor}
\pgfsetstrokeopacity{1.000000}
\draw (11.000000\du,-8.000000\du)--(11.000000\du,-6.000000\du)--(17.000000\du,-6.000000\du)--(17.000000\du,-8.000000\du)--cycle;
}
\definecolor{dialinecolor}{rgb}{0.000000, 0.000000, 0.000000}
\pgfsetstrokecolor{dialinecolor}
\pgfsetstrokeopacity{1.000000}
\definecolor{diafillcolor}{rgb}{0.000000, 0.000000, 0.000000}
\pgfsetfillcolor{diafillcolor}
\pgfsetfillopacity{1.000000}
\node[anchor=base,inner sep=0pt, outer sep=0pt,color=dialinecolor] at (14.000000\du,-6.805000\du){LSTM};
\pgfsetlinewidth{0.100000\du}
\pgfsetdash{}{0pt}
\pgfsetmiterjoin
{\pgfsetcornersarced{\pgfpoint{0.000000\du}{0.000000\du}}\definecolor{diafillcolor}{rgb}{0.000000, 1.000000, 0.000000}
\pgfsetfillcolor{diafillcolor}
\pgfsetfillopacity{1.000000}
\fill (19.000000\du,14.000000\du)--(19.000000\du,16.050000\du)--(25.000000\du,16.050000\du)--(25.000000\du,14.000000\du)--cycle;
}{\pgfsetcornersarced{\pgfpoint{0.000000\du}{0.000000\du}}\definecolor{dialinecolor}{rgb}{0.000000, 0.000000, 0.000000}
\pgfsetstrokecolor{dialinecolor}
\pgfsetstrokeopacity{1.000000}
\draw (19.000000\du,14.000000\du)--(19.000000\du,16.050000\du)--(25.000000\du,16.050000\du)--(25.000000\du,14.000000\du)--cycle;
}
\definecolor{dialinecolor}{rgb}{0.000000, 0.000000, 0.000000}
\pgfsetstrokecolor{dialinecolor}
\pgfsetstrokeopacity{1.000000}
\definecolor{diafillcolor}{rgb}{0.000000, 0.000000, 0.000000}
\pgfsetfillcolor{diafillcolor}
\pgfsetfillopacity{1.000000}
\node[anchor=base,inner sep=0pt, outer sep=0pt,color=dialinecolor] at (22.000000\du,15.220000\du){LSTM};
\pgfsetlinewidth{0.100000\du}
\pgfsetdash{}{0pt}
\pgfsetmiterjoin
{\pgfsetcornersarced{\pgfpoint{0.000000\du}{0.000000\du}}\definecolor{diafillcolor}{rgb}{1.000000, 0.647059, 0.000000}
\pgfsetfillcolor{diafillcolor}
\pgfsetfillopacity{1.000000}
\fill (19.000000\du,11.000000\du)--(19.000000\du,13.050000\du)--(25.000000\du,13.050000\du)--(25.000000\du,11.000000\du)--cycle;
}{\pgfsetcornersarced{\pgfpoint{0.000000\du}{0.000000\du}}\definecolor{dialinecolor}{rgb}{0.000000, 0.000000, 0.000000}
\pgfsetstrokecolor{dialinecolor}
\pgfsetstrokeopacity{1.000000}
\draw (19.000000\du,11.000000\du)--(19.000000\du,13.050000\du)--(25.000000\du,13.050000\du)--(25.000000\du,11.000000\du)--cycle;
}
\definecolor{dialinecolor}{rgb}{0.000000, 0.000000, 0.000000}
\pgfsetstrokecolor{dialinecolor}
\pgfsetstrokeopacity{1.000000}
\definecolor{diafillcolor}{rgb}{0.000000, 0.000000, 0.000000}
\pgfsetfillcolor{diafillcolor}
\pgfsetfillopacity{1.000000}
\node[anchor=base,inner sep=0pt, outer sep=0pt,color=dialinecolor] at (22.000000\du,12.220000\du){MAXPOOL};
\pgfsetlinewidth{0.100000\du}
\pgfsetdash{}{0pt}
\pgfsetmiterjoin
{\pgfsetcornersarced{\pgfpoint{0.000000\du}{0.000000\du}}\definecolor{diafillcolor}{rgb}{0.117647, 0.564706, 1.000000}
\pgfsetfillcolor{diafillcolor}
\pgfsetfillopacity{1.000000}
\fill (11.000000\du,-15.000000\du)--(11.000000\du,-12.000000\du)--(17.000000\du,-12.000000\du)--(17.000000\du,-15.000000\du)--cycle;
}{\pgfsetcornersarced{\pgfpoint{0.000000\du}{0.000000\du}}\definecolor{dialinecolor}{rgb}{0.000000, 0.000000, 0.000000}
\pgfsetstrokecolor{dialinecolor}
\pgfsetstrokeopacity{1.000000}
\draw (11.000000\du,-15.000000\du)--(11.000000\du,-12.000000\du)--(17.000000\du,-12.000000\du)--(17.000000\du,-15.000000\du)--cycle;
}
\definecolor{dialinecolor}{rgb}{0.000000, 0.000000, 0.000000}
\pgfsetstrokecolor{dialinecolor}
\pgfsetstrokeopacity{1.000000}
\definecolor{diafillcolor}{rgb}{0.000000, 0.000000, 0.000000}
\pgfsetfillcolor{diafillcolor}
\pgfsetfillopacity{1.000000}
\node[anchor=base,inner sep=0pt, outer sep=0pt,color=dialinecolor] at (14.000000\du,-13.705000\du){SOFTMAX};
\definecolor{dialinecolor}{rgb}{0.000000, 0.000000, 0.000000}
\pgfsetstrokecolor{dialinecolor}
\pgfsetstrokeopacity{1.000000}
\definecolor{diafillcolor}{rgb}{0.000000, 0.000000, 0.000000}
\pgfsetfillcolor{diafillcolor}
\pgfsetfillopacity{1.000000}
\node[anchor=base,inner sep=0pt, outer sep=0pt,color=dialinecolor] at (14.000000\du,-12.905000\du){(MLP\_readout)};
\pgfsetlinewidth{0.100000\du}
\pgfsetdash{}{0pt}
\pgfsetbuttcap
{
\definecolor{diafillcolor}{rgb}{0.000000, 0.000000, 0.000000}
\pgfsetfillcolor{diafillcolor}
\pgfsetfillopacity{1.000000}
\pgfsetarrowsend{stealth}
\definecolor{dialinecolor}{rgb}{0.000000, 0.000000, 0.000000}
\pgfsetstrokecolor{dialinecolor}
\pgfsetstrokeopacity{1.000000}
\draw (22.000000\du,24.000000\du)--(22.000000\du,22.050000\du);
}
\pgfsetlinewidth{0.100000\du}
\pgfsetdash{}{0pt}
\pgfsetbuttcap
{
\definecolor{diafillcolor}{rgb}{0.000000, 0.000000, 0.000000}
\pgfsetfillcolor{diafillcolor}
\pgfsetfillopacity{1.000000}
\pgfsetarrowsend{stealth}
\definecolor{dialinecolor}{rgb}{0.000000, 0.000000, 0.000000}
\pgfsetstrokecolor{dialinecolor}
\pgfsetstrokeopacity{1.000000}
\draw (22.000000\du,20.000000\du)--(22.000000\du,19.050000\du);
}
\pgfsetlinewidth{0.100000\du}
\pgfsetdash{}{0pt}
\pgfsetbuttcap
{
\definecolor{diafillcolor}{rgb}{0.000000, 0.000000, 0.000000}
\pgfsetfillcolor{diafillcolor}
\pgfsetfillopacity{1.000000}
\pgfsetarrowsend{stealth}
\definecolor{dialinecolor}{rgb}{0.000000, 0.000000, 0.000000}
\pgfsetstrokecolor{dialinecolor}
\pgfsetstrokeopacity{1.000000}
\draw (22.000000\du,17.000000\du)--(22.000000\du,16.050000\du);
}
\pgfsetlinewidth{0.100000\du}
\pgfsetdash{}{0pt}
\pgfsetbuttcap
{
\definecolor{diafillcolor}{rgb}{0.000000, 0.000000, 0.000000}
\pgfsetfillcolor{diafillcolor}
\pgfsetfillopacity{1.000000}
\pgfsetarrowsend{stealth}
\definecolor{dialinecolor}{rgb}{0.000000, 0.000000, 0.000000}
\pgfsetstrokecolor{dialinecolor}
\pgfsetstrokeopacity{1.000000}
\draw (22.000000\du,14.000000\du)--(22.000000\du,13.050000\du);
}
\definecolor{dialinecolor}{rgb}{0.000000, 0.000000, 0.000000}
\pgfsetstrokecolor{dialinecolor}
\pgfsetstrokeopacity{1.000000}
\definecolor{diafillcolor}{rgb}{0.000000, 0.000000, 0.000000}
\pgfsetfillcolor{diafillcolor}
\pgfsetfillopacity{1.000000}
\node[anchor=base west,inner sep=0pt,outer sep=0pt,color=dialinecolor] at (21.000000\du,6.000000\du){};
\definecolor{dialinecolor}{rgb}{0.000000, 0.000000, 0.000000}
\pgfsetstrokecolor{dialinecolor}
\pgfsetstrokeopacity{1.000000}
\definecolor{diafillcolor}{rgb}{0.000000, 0.000000, 0.000000}
\pgfsetfillcolor{diafillcolor}
\pgfsetfillopacity{1.000000}
\node[anchor=base west,inner sep=0pt,outer sep=0pt,color=dialinecolor] at
(21.9000\du,8.400000\du){.};
\definecolor{dialinecolor}{rgb}{0.000000, 0.000000, 0.000000}
\pgfsetstrokecolor{dialinecolor}
\pgfsetstrokeopacity{1.000000}
\definecolor{diafillcolor}{rgb}{0.000000, 0.000000, 0.000000}
\pgfsetfillcolor{diafillcolor}
\pgfsetfillopacity{1.000000}
\node[anchor=base west,inner sep=0pt,outer sep=0pt,color=dialinecolor] at 
(21.9000\du,9.000000\du){.};
\definecolor{dialinecolor}{rgb}{0.000000, 0.000000, 0.000000}
\pgfsetstrokecolor{dialinecolor}
\pgfsetstrokeopacity{1.000000}
\definecolor{diafillcolor}{rgb}{0.000000, 0.000000, 0.000000}
\pgfsetfillcolor{diafillcolor}
\pgfsetfillopacity{1.000000}
\node[anchor=base west,inner sep=0pt,outer sep=0pt,color=dialinecolor] at 
(21.9000\du,9.600000\du){.};
\pgfsetlinewidth{0.100000\du}
\pgfsetdash{}{0pt}
\pgfsetbuttcap
{
\definecolor{diafillcolor}{rgb}{0.000000, 0.000000, 0.000000}
\pgfsetfillcolor{diafillcolor}
\pgfsetfillopacity{1.000000}
\pgfsetarrowsend{stealth}
\definecolor{dialinecolor}{rgb}{0.000000, 0.000000, 0.000000}
\pgfsetstrokecolor{dialinecolor}
\pgfsetstrokeopacity{1.000000}
\draw (22.000000\du,11.000000\du)--(22.000000\du,10.050000\du);
}
\pgfsetlinewidth{0.100000\du}
\pgfsetdash{}{0pt}
\pgfsetbuttcap
{
\definecolor{diafillcolor}{rgb}{0.000000, 0.000000, 0.000000}
\pgfsetfillcolor{diafillcolor}
\pgfsetfillopacity{1.000000}
\pgfsetarrowsend{stealth}
\definecolor{dialinecolor}{rgb}{0.000000, 0.000000, 0.000000}
\pgfsetstrokecolor{dialinecolor}
\pgfsetstrokeopacity{1.000000}
\draw (22.000000\du,8.000000\du)--(22.000000\du,7.050000\du);
}
\pgfsetlinewidth{0.100000\du}
\pgfsetdash{}{0pt}
\pgfsetmiterjoin
\definecolor{diafillcolor}{rgb}{1.000000, 1.000000, 1.000000}
\pgfsetfillcolor{diafillcolor}
\pgfsetfillopacity{1.000000}
\pgfpathellipse{\pgfpoint{25.883243\du}{21.050000\du}}{\pgfpoint{1.258243\du}{0\du}}{\pgfpoint{0\du}{1.250000\du}}
\pgfusepath{fill}
\definecolor{dialinecolor}{rgb}{0.000000, 0.000000, 0.000000}
\pgfsetstrokecolor{dialinecolor}
\pgfsetstrokeopacity{1.000000}
\pgfpathellipse{\pgfpoint{25.883243\du}{21.050000\du}}{\pgfpoint{1.258243\du}{0\du}}{\pgfpoint{0\du}{1.250000\du}}
\pgfusepath{stroke}
\definecolor{dialinecolor}{rgb}{0.000000, 0.000000, 0.000000}
\pgfsetstrokecolor{dialinecolor}
\pgfsetstrokeopacity{1.000000}
\definecolor{diafillcolor}{rgb}{0.000000, 0.000000, 0.000000}
\pgfsetfillcolor{diafillcolor}
\pgfsetfillopacity{1.000000}
\node[anchor=base,inner sep=0pt, outer sep=0pt,color=dialinecolor] at (25.883243\du,21.245000\du){};
\pgfsetlinewidth{0.100000\du}
\pgfsetdash{}{0pt}
\pgfsetmiterjoin
{\pgfsetcornersarced{\pgfpoint{0.000000\du}{0.000000\du}}\definecolor{diafillcolor}{rgb}{0.000000, 1.000000, 0.000000}
\pgfsetfillcolor{diafillcolor}
\pgfsetfillopacity{1.000000}
\fill (19.000000\du,20.000000\du)--(19.000000\du,22.050000\du)--(25.000000\du,22.050000\du)--(25.000000\du,20.000000\du)--cycle;
}{\pgfsetcornersarced{\pgfpoint{0.000000\du}{0.000000\du}}\definecolor{dialinecolor}{rgb}{0.000000, 0.000000, 0.000000}
\pgfsetstrokecolor{dialinecolor}
\pgfsetstrokeopacity{1.000000}
\draw (19.000000\du,20.000000\du)--(19.000000\du,22.050000\du)--(25.000000\du,22.050000\du)--(25.000000\du,20.000000\du)--cycle;
}
\definecolor{dialinecolor}{rgb}{0.000000, 0.000000, 0.000000}
\pgfsetstrokecolor{dialinecolor}
\pgfsetstrokeopacity{1.000000}
\definecolor{diafillcolor}{rgb}{0.000000, 0.000000, 0.000000}
\pgfsetfillcolor{diafillcolor}
\pgfsetfillopacity{1.000000}
\node[anchor=base,inner sep=0pt, outer sep=0pt,color=dialinecolor] at (22.000000\du,21.220000\du){LSTM};
\pgfsetlinewidth{0.100000\du}
\pgfsetdash{}{0pt}
\pgfsetbuttcap
{
\definecolor{diafillcolor}{rgb}{0.000000, 0.000000, 0.000000}
\pgfsetfillcolor{diafillcolor}
\pgfsetfillopacity{1.000000}
\pgfsetarrowsend{stealth}
\definecolor{dialinecolor}{rgb}{0.000000, 0.000000, 0.000000}
\pgfsetstrokecolor{dialinecolor}
\pgfsetstrokeopacity{1.000000}
\draw (25.462200\du,22.246200\du)--(25.062200\du,21.946200\du);
}
\pgfsetlinewidth{0.100000\du}
\pgfsetdash{}{0pt}
\pgfsetbuttcap
{
\definecolor{diafillcolor}{rgb}{0.000000, 0.000000, 0.000000}
\pgfsetfillcolor{diafillcolor}
\pgfsetfillopacity{1.000000}
\pgfsetarrowsend{stealth}
\definecolor{dialinecolor}{rgb}{0.000000, 0.000000, 0.000000}
\pgfsetstrokecolor{dialinecolor}
\pgfsetstrokeopacity{1.000000}
\draw (25.368300\du,16.115700\du)--(24.960100\du,15.844700\du);
}
\pgfsetlinewidth{0.100000\du}
\pgfsetdash{}{0pt}
\pgfsetbuttcap
{
\definecolor{diafillcolor}{rgb}{0.000000, 0.000000, 0.000000}
\pgfsetfillcolor{diafillcolor}
\pgfsetfillopacity{1.000000}
\pgfsetarrowsend{stealth}
\definecolor{dialinecolor}{rgb}{0.000000, 0.000000, 0.000000}
\pgfsetstrokecolor{dialinecolor}
\pgfsetstrokeopacity{1.000000}
\draw (25.390300\du,7.104960\du)--(24.982100\du,6.834000\du);
}
\pgfsetlinewidth{0.100000\du}
\pgfsetdash{}{0pt}
\pgfsetbuttcap
{
\definecolor{diafillcolor}{rgb}{0.000000, 0.000000, 0.000000}
\pgfsetfillcolor{diafillcolor}
\pgfsetfillopacity{1.000000}
\pgfsetarrowsend{stealth}
\definecolor{dialinecolor}{rgb}{0.000000, 0.000000, 0.000000}
\pgfsetstrokecolor{dialinecolor}
\pgfsetstrokeopacity{1.000000}
\draw (22.000000\du,5.000000\du)--(22.000000\du,1.000000\du);
}
\pgfsetlinewidth{0.100000\du}
\pgfsetdash{}{0pt}
\pgfsetbuttcap
{
\definecolor{diafillcolor}{rgb}{0.000000, 0.000000, 0.000000}
\pgfsetfillcolor{diafillcolor}
\pgfsetfillopacity{1.000000}
\pgfsetarrowsend{stealth}
\definecolor{dialinecolor}{rgb}{0.000000, 0.000000, 0.000000}
\pgfsetstrokecolor{dialinecolor}
\pgfsetstrokeopacity{1.000000}
\draw (16.000000\du,-9.000000\du)--(16.000000\du,-12.000000\du);
}
\pgfsetlinewidth{0.100000\du}
\pgfsetdash{}{0pt}
\pgfsetbuttcap
{
\definecolor{diafillcolor}{rgb}{0.000000, 0.000000, 0.000000}
\pgfsetfillcolor{diafillcolor}
\pgfsetfillopacity{1.000000}
\pgfsetarrowsend{stealth}
\definecolor{dialinecolor}{rgb}{0.000000, 0.000000, 0.000000}
\pgfsetstrokecolor{dialinecolor}
\pgfsetstrokeopacity{1.000000}
\draw (14.000000\du,-8.000000\du)--(14.000000\du,-12.000000\du);
}
\pgfsetlinewidth{0.100000\du}
\pgfsetdash{}{0pt}
\pgfsetbuttcap
{
\definecolor{diafillcolor}{rgb}{0.000000, 0.000000, 0.000000}
\pgfsetfillcolor{diafillcolor}
\pgfsetfillopacity{1.000000}
\pgfsetarrowsend{stealth}
\definecolor{dialinecolor}{rgb}{0.000000, 0.000000, 0.000000}
\pgfsetstrokecolor{dialinecolor}
\pgfsetstrokeopacity{1.000000}
\draw (14.000000\du,-15.000000\du)--(14.000000\du,-16.000000\du);
}
\pgfsetlinewidth{0.100000\du}
\pgfsetdash{}{0pt}
\pgfsetbuttcap
{
\definecolor{diafillcolor}{rgb}{0.000000, 0.000000, 0.000000}
\pgfsetfillcolor{diafillcolor}
\pgfsetfillopacity{1.000000}
\pgfsetarrowsend{stealth}
\definecolor{dialinecolor}{rgb}{0.000000, 0.000000, 0.000000}
\pgfsetstrokecolor{dialinecolor}
\pgfsetstrokeopacity{1.000000}
\draw (17.432400\du,-5.801970\du)--(17.024200\du,-6.072940\du);
}
\pgfsetlinewidth{0.100000\du}
\pgfsetdash{}{0pt}
\pgfsetbuttcap
{
\definecolor{diafillcolor}{rgb}{0.000000, 0.000000, 0.000000}
\pgfsetfillcolor{diafillcolor}
\pgfsetfillopacity{1.000000}
\definecolor{dialinecolor}{rgb}{0.000000, 0.000000, 0.000000}
\pgfsetstrokecolor{dialinecolor}
\pgfsetstrokeopacity{1.000000}
\draw (28.000000\du,5.000000\du)--(30.150000\du,14.000000\du);
}
\pgfsetlinewidth{0.100000\du}
\pgfsetdash{}{0pt}
\pgfsetbuttcap
{
\definecolor{diafillcolor}{rgb}{0.000000, 0.000000, 0.000000}
\pgfsetfillcolor{diafillcolor}
\pgfsetfillopacity{1.000000}
\definecolor{dialinecolor}{rgb}{0.000000, 0.000000, 0.000000}
\pgfsetstrokecolor{dialinecolor}
\pgfsetstrokeopacity{1.000000}
\draw (28.150000\du,22.850000\du)--(30.150000\du,14.000000\du);
}
\definecolor{dialinecolor}{rgb}{0.000000, 0.000000, 0.000000}
\pgfsetstrokecolor{dialinecolor}
\pgfsetstrokeopacity{1.000000}
\definecolor{diafillcolor}{rgb}{0.000000, 0.000000, 0.000000}
\pgfsetfillcolor{diafillcolor}
\pgfsetfillopacity{1.000000}
\node[anchor=base west,inner sep=0pt,outer sep=0pt,color=dialinecolor] at
(22.257900\du,23.354300\du){$\vec{x}_{\text{sem}}[m]$};
\definecolor{dialinecolor}{rgb}{0.000000, 0.000000, 0.000000}
\pgfsetstrokecolor{dialinecolor}
\pgfsetstrokeopacity{1.000000}
\definecolor{diafillcolor}{rgb}{0.000000, 0.000000, 0.000000}
\pgfsetfillcolor{diafillcolor}
\pgfsetfillopacity{1.000000}
\node[anchor=base west,inner sep=0pt,outer sep=0pt,color=dialinecolor] at (31.150000\du,16.000000\du){Encoder};
\definecolor{dialinecolor}{rgb}{0.000000, 0.000000, 0.000000}
\pgfsetstrokecolor{dialinecolor}
\pgfsetstrokeopacity{1.000000}
\definecolor{diafillcolor}{rgb}{0.000000, 0.000000, 0.000000}
\pgfsetfillcolor{diafillcolor}
\pgfsetfillopacity{1.000000}
\node[anchor=base west,inner sep=0pt,outer sep=0pt,color=dialinecolor] at
(15.000000\du,-19.000000\du){$\hat{y}_l$};
\pgfsetlinewidth{0.100000\du}
\pgfsetdash{}{0pt}
\pgfsetbuttcap
{
\definecolor{diafillcolor}{rgb}{0.000000, 0.000000, 0.000000}
\pgfsetfillcolor{diafillcolor}
\pgfsetfillopacity{1.000000}
\definecolor{dialinecolor}{rgb}{0.000000, 0.000000, 0.000000}
\pgfsetstrokecolor{dialinecolor}
\pgfsetstrokeopacity{1.000000}
\draw (18.000000\du,-2.000000\du)--(15.000000\du,-2.000000\du);
}
\pgfsetlinewidth{0.100000\du}
\pgfsetdash{}{0pt}
\pgfsetbuttcap
{
\definecolor{diafillcolor}{rgb}{0.000000, 0.000000, 0.000000}
\pgfsetfillcolor{diafillcolor}
\pgfsetfillopacity{1.000000}
\pgfsetarrowsend{stealth}
\definecolor{dialinecolor}{rgb}{0.000000, 0.000000, 0.000000}
\pgfsetstrokecolor{dialinecolor}
\pgfsetstrokeopacity{1.000000}
\draw (20.000000\du,3.000000\du)--(20.000000\du,1.000000\du);
}
\pgfsetlinewidth{0.100000\du}
\pgfsetdash{}{0pt}
\pgfsetbuttcap
{
\definecolor{diafillcolor}{rgb}{0.000000, 0.000000, 0.000000}
\pgfsetfillcolor{diafillcolor}
\pgfsetfillopacity{1.000000}
\definecolor{dialinecolor}{rgb}{0.000000, 0.000000, 0.000000}
\pgfsetstrokecolor{dialinecolor}
\pgfsetstrokeopacity{1.000000}
\draw (14.000000\du,-9.000000\du)--(8.000000\du,-9.000000\du);
}
\pgfsetlinewidth{0.100000\du}
\pgfsetdash{}{0pt}
\pgfsetbuttcap
{
\definecolor{diafillcolor}{rgb}{0.000000, 0.000000, 0.000000}
\pgfsetfillcolor{diafillcolor}
\pgfsetfillopacity{1.000000}
\definecolor{dialinecolor}{rgb}{0.000000, 0.000000, 0.000000}
\pgfsetstrokecolor{dialinecolor}
\pgfsetstrokeopacity{1.000000}
\draw (14.000000\du,-19.000000\du)--(5.021250\du,-19.000000\du);
}
\pgfsetlinewidth{0.100000\du}
\pgfsetdash{}{0pt}
\pgfsetbuttcap
{
\definecolor{diafillcolor}{rgb}{0.000000, 0.000000, 0.000000}
\pgfsetfillcolor{diafillcolor}
\pgfsetfillopacity{1.000000}
\definecolor{dialinecolor}{rgb}{0.000000, 0.000000, 0.000000}
\pgfsetstrokecolor{dialinecolor}
\pgfsetstrokeopacity{1.000000}
\draw (8.000000\du,3.000000\du)--(8.000000\du,-9.000000\du);
}
\pgfsetlinewidth{0.100000\du}
\pgfsetdash{}{0pt}
\pgfsetbuttcap
{
\definecolor{diafillcolor}{rgb}{0.000000, 0.000000, 0.000000}
\pgfsetfillcolor{diafillcolor}
\pgfsetfillopacity{1.000000}
\definecolor{dialinecolor}{rgb}{0.000000, 0.000000, 0.000000}
\pgfsetstrokecolor{dialinecolor}
\pgfsetstrokeopacity{1.000000}
\draw (20.000000\du,3.000000\du)--(8.000000\du,3.000000\du);
}
\pgfsetlinewidth{0.100000\du}
\pgfsetdash{}{0pt}
\pgfsetbuttcap
{
\definecolor{diafillcolor}{rgb}{0.000000, 0.000000, 0.000000}
\pgfsetfillcolor{diafillcolor}
\pgfsetfillopacity{1.000000}
\definecolor{dialinecolor}{rgb}{0.000000, 0.000000, 0.000000}
\pgfsetstrokecolor{dialinecolor}
\pgfsetstrokeopacity{1.000000}
\draw (20.000000\du,-1.000000\du)--(20.000000\du,-9.000000\du);
}
\pgfsetlinewidth{0.100000\du}
\pgfsetdash{}{0pt}
\pgfsetbuttcap
{
\definecolor{diafillcolor}{rgb}{0.000000, 0.000000, 0.000000}
\pgfsetfillcolor{diafillcolor}
\pgfsetfillopacity{1.000000}
\definecolor{dialinecolor}{rgb}{0.000000, 0.000000, 0.000000}
\pgfsetstrokecolor{dialinecolor}
\pgfsetstrokeopacity{1.000000}
\draw (28.000000\du,-8.000000\du)--(30.000000\du,-7.000000\du);
}
\pgfsetlinewidth{0.100000\du}
\pgfsetdash{}{0pt}
\pgfsetbuttcap
{
\definecolor{diafillcolor}{rgb}{0.000000, 0.000000, 0.000000}
\pgfsetfillcolor{diafillcolor}
\pgfsetfillopacity{1.000000}
\definecolor{dialinecolor}{rgb}{0.000000, 0.000000, 0.000000}
\pgfsetstrokecolor{dialinecolor}
\pgfsetstrokeopacity{1.000000}
\draw (30.000000\du,-7.000000\du)--(28.000000\du,-6.000000\du);
}
\definecolor{dialinecolor}{rgb}{0.000000, 0.000000, 0.000000}
\pgfsetstrokecolor{dialinecolor}
\pgfsetstrokeopacity{1.000000}
\definecolor{diafillcolor}{rgb}{0.000000, 0.000000, 0.000000}
\pgfsetfillcolor{diafillcolor}
\pgfsetfillopacity{1.000000}
\node[anchor=base west,inner sep=0pt,outer sep=0pt,color=dialinecolor] at (31.000000\du,-7.000000\du){Decoder};
\pgfsetlinewidth{0.100000\du}
\pgfsetdash{}{0pt}
\pgfsetbuttcap
{
\definecolor{diafillcolor}{rgb}{0.000000, 0.000000, 0.000000}
\pgfsetfillcolor{diafillcolor}
\pgfsetfillopacity{1.000000}
\definecolor{dialinecolor}{rgb}{0.000000, 0.000000, 0.000000}
\pgfsetstrokecolor{dialinecolor}
\pgfsetstrokeopacity{1.000000}
\draw (16.000000\du,-9.000000\du)--(20.000000\du,-9.000000\du);
}
\pgfsetlinewidth{0.100000\du}
\pgfsetdash{}{0pt}
\pgfsetbuttcap
{
\definecolor{diafillcolor}{rgb}{0.000000, 0.000000, 0.000000}
\pgfsetfillcolor{diafillcolor}
\pgfsetfillopacity{1.000000}
\definecolor{dialinecolor}{rgb}{0.000000, 0.000000, 0.000000}
\pgfsetstrokecolor{dialinecolor}
\pgfsetstrokeopacity{1.000000}
\draw (18.000000\du,-1.000000\du)--(18.000000\du,-2.000000\du);
}
\pgfsetlinewidth{0.100000\du}
\pgfsetdash{}{0pt}
\pgfsetbuttcap
{
\definecolor{diafillcolor}{rgb}{0.000000, 0.000000, 0.000000}
\pgfsetfillcolor{diafillcolor}
\pgfsetfillopacity{1.000000}
\definecolor{dialinecolor}{rgb}{0.000000, 0.000000, 0.000000}
\pgfsetstrokecolor{dialinecolor}
\pgfsetstrokeopacity{1.000000}
\draw (15.000000\du,-2.000000\du)--(15.000000\du,2.054550\du);
}
\pgfsetlinewidth{0.100000\du}
\pgfsetdash{}{0pt}
\pgfsetbuttcap
{
\definecolor{diafillcolor}{rgb}{0.000000, 0.000000, 0.000000}
\pgfsetfillcolor{diafillcolor}
\pgfsetfillopacity{1.000000}
\definecolor{dialinecolor}{rgb}{0.000000, 0.000000, 0.000000}
\pgfsetstrokecolor{dialinecolor}
\pgfsetstrokeopacity{1.000000}
\draw (15.000000\du,2.000000\du)--(18.000000\du,2.000000\du);
}
\pgfsetlinewidth{0.100000\du}
\pgfsetdash{}{0pt}
\pgfsetbuttcap
{
\definecolor{diafillcolor}{rgb}{0.000000, 0.000000, 0.000000}
\pgfsetfillcolor{diafillcolor}
\pgfsetfillopacity{1.000000}
\pgfsetarrowsend{stealth}
\definecolor{dialinecolor}{rgb}{0.000000, 0.000000, 0.000000}
\pgfsetstrokecolor{dialinecolor}
\pgfsetstrokeopacity{1.000000}
\draw (18.000000\du,2.000000\du)--(18.000000\du,1.000000\du);
}
\pgfsetlinewidth{0.100000\du}
\pgfsetdash{}{0pt}
\pgfsetbuttcap
{
\definecolor{diafillcolor}{rgb}{0.000000, 0.000000, 0.000000}
\pgfsetfillcolor{diafillcolor}
\pgfsetfillopacity{1.000000}
\pgfsetarrowsend{stealth}
\definecolor{dialinecolor}{rgb}{0.000000, 0.000000, 0.000000}
\pgfsetstrokecolor{dialinecolor}
\pgfsetstrokeopacity{1.000000}
\draw (5.000000\du,-19.000000\du)--(4.957500\du,-16.000000\du);
}
\pgfsetlinewidth{0.100000\du}
\pgfsetdash{}{0pt}
\pgfsetbuttcap
{
\definecolor{diafillcolor}{rgb}{0.000000, 0.000000, 0.000000}
\pgfsetfillcolor{diafillcolor}
\pgfsetfillopacity{1.000000}
\pgfsetarrowsstart{stealth}
\definecolor{dialinecolor}{rgb}{0.000000, 0.000000, 0.000000}
\pgfsetstrokecolor{dialinecolor}
\pgfsetstrokeopacity{1.000000}
\draw (12.000000\du,-12.000000\du)--(12.000000\du,-11.000000\du);
}
\pgfsetlinewidth{0.100000\du}
\pgfsetdash{}{0pt}
\pgfsetbuttcap
{
\definecolor{diafillcolor}{rgb}{0.000000, 0.000000, 0.000000}
\pgfsetfillcolor{diafillcolor}
\pgfsetfillopacity{1.000000}
\definecolor{dialinecolor}{rgb}{0.000000, 0.000000, 0.000000}
\pgfsetstrokecolor{dialinecolor}
\pgfsetstrokeopacity{1.000000}
\draw (12.000000\du,-11.000000\du)--(5.000000\du,-11.000000\du);
}
\pgfsetlinewidth{0.100000\du}
\pgfsetdash{}{0pt}
\pgfsetbuttcap
{
\definecolor{diafillcolor}{rgb}{0.000000, 0.000000, 0.000000}
\pgfsetfillcolor{diafillcolor}
\pgfsetfillopacity{1.000000}
\definecolor{dialinecolor}{rgb}{0.000000, 0.000000, 0.000000}
\pgfsetstrokecolor{dialinecolor}
\pgfsetstrokeopacity{1.000000}
\draw (4.957500\du,-13.950000\du)--(5.000000\du,-4.000000\du);
}
\pgfsetlinewidth{0.100000\du}
\pgfsetdash{}{0pt}
\pgfsetbuttcap
{
\definecolor{diafillcolor}{rgb}{0.000000, 0.000000, 0.000000}
\pgfsetfillcolor{diafillcolor}
\pgfsetfillopacity{1.000000}
\definecolor{dialinecolor}{rgb}{0.000000, 0.000000, 0.000000}
\pgfsetstrokecolor{dialinecolor}
\pgfsetstrokeopacity{1.000000}
\draw (13.000000\du,-4.000000\du)--(5.000000\du,-4.000000\du);
}
\pgfsetlinewidth{0.100000\du}
\pgfsetdash{}{0pt}
\pgfsetbuttcap
{
\definecolor{diafillcolor}{rgb}{0.000000, 0.000000, 0.000000}
\pgfsetfillcolor{diafillcolor}
\pgfsetfillopacity{1.000000}
\pgfsetarrowsend{stealth}
\definecolor{dialinecolor}{rgb}{0.000000, 0.000000, 0.000000}
\pgfsetstrokecolor{dialinecolor}
\pgfsetstrokeopacity{1.000000}
\draw (13.000000\du,-4.000000\du)--(13.000000\du,-6.000000\du);
}
\pgfsetlinewidth{0.100000\du}
\pgfsetdash{}{0pt}
\pgfsetbuttcap
{
\definecolor{diafillcolor}{rgb}{0.000000, 0.000000, 0.000000}
\pgfsetfillcolor{diafillcolor}
\pgfsetfillopacity{1.000000}
\definecolor{dialinecolor}{rgb}{0.000000, 0.000000, 0.000000}
\pgfsetstrokecolor{dialinecolor}
\pgfsetstrokeopacity{1.000000}
\draw (15.000000\du,-4.000000\du)--(20.000000\du,-4.000000\du);
}
\pgfsetlinewidth{0.100000\du}
\pgfsetdash{}{0pt}
\pgfsetbuttcap
{
\definecolor{diafillcolor}{rgb}{0.000000, 0.000000, 0.000000}
\pgfsetfillcolor{diafillcolor}
\pgfsetfillopacity{1.000000}
\pgfsetarrowsend{stealth}
\definecolor{dialinecolor}{rgb}{0.000000, 0.000000, 0.000000}
\pgfsetstrokecolor{dialinecolor}
\pgfsetstrokeopacity{1.000000}
\draw (15.000000\du,-4.000000\du)--(15.000000\du,-6.054550\du);
}
\definecolor{dialinecolor}{rgb}{0.000000, 0.000000, 0.000000}
\pgfsetstrokecolor{dialinecolor}
\pgfsetstrokeopacity{1.000000}
\definecolor{diafillcolor}{rgb}{0.000000, 0.000000, 0.000000}
\pgfsetfillcolor{diafillcolor}
\pgfsetfillopacity{1.000000}
\node[anchor=base west,inner sep=0pt,outer sep=0pt,color=dialinecolor] at
(17.000000\du,-11.000000\du){$\vec{c}_l$};
\definecolor{dialinecolor}{rgb}{0.000000, 0.000000, 0.000000}
\pgfsetstrokecolor{dialinecolor}
\pgfsetstrokeopacity{1.000000}
\definecolor{diafillcolor}{rgb}{0.000000, 0.000000, 0.000000}
\pgfsetfillcolor{diafillcolor}
\pgfsetfillopacity{1.000000}
\node[anchor=base west,inner sep=0pt,outer sep=0pt,color=dialinecolor] at
(7.000000\du,-10.000000\du){$\vec{y}^{\,\,e}_{l-1}$};
\definecolor{dialinecolor}{rgb}{0.000000, 0.000000, 0.000000}
\pgfsetstrokecolor{dialinecolor}
\pgfsetstrokeopacity{1.000000}
\definecolor{diafillcolor}{rgb}{0.000000, 0.000000, 0.000000}
\pgfsetfillcolor{diafillcolor}
\pgfsetfillopacity{1.000000}
\node[anchor=base west,inner sep=0pt,outer sep=0pt,color=dialinecolor] at
(15.000000\du,-3.000000\du){$\vec{c}_l$};
\definecolor{dialinecolor}{rgb}{0.000000, 0.000000, 0.000000}
\pgfsetstrokecolor{dialinecolor}
\pgfsetstrokeopacity{1.000000}
\definecolor{diafillcolor}{rgb}{0.000000, 0.000000, 0.000000}
\pgfsetfillcolor{diafillcolor}
\pgfsetfillopacity{1.000000}
\node[anchor=base west,inner sep=0pt,outer sep=0pt,color=dialinecolor] at
(10.000000\du,-3.000000\du){$\vec{y}^{\,\,e}_{l-1}$};
\definecolor{dialinecolor}{rgb}{0.000000, 0.000000, 0.000000}
\pgfsetstrokecolor{dialinecolor}
\pgfsetstrokeopacity{1.000000}
\definecolor{diafillcolor}{rgb}{0.000000, 0.000000, 0.000000}
\pgfsetfillcolor{diafillcolor}
\pgfsetfillopacity{1.000000}
\node[anchor=base west,inner sep=0pt,outer sep=0pt,color=dialinecolor] at
(12.800000\du,0.000000\du){$\vec{\beta}_l[m]$};
\definecolor{dialinecolor}{rgb}{0.000000, 0.000000, 0.000000}
\pgfsetstrokecolor{dialinecolor}
\pgfsetstrokeopacity{1.000000}
\definecolor{diafillcolor}{rgb}{0.000000, 0.000000, 0.000000}
\pgfsetfillcolor{diafillcolor}
\pgfsetfillopacity{1.000000}
\node[anchor=base west,inner sep=0pt,outer sep=0pt,color=dialinecolor] at
(22.600000\du,3.700000\du)
{$\vec{h}^{enc}[m]$};
\definecolor{dialinecolor}{rgb}{0.000000, 0.000000, 0.000000}
\pgfsetstrokecolor{dialinecolor}
\pgfsetstrokeopacity{1.000000}
\definecolor{diafillcolor}{rgb}{0.000000, 0.000000, 0.000000}
\pgfsetfillcolor{diafillcolor}
\pgfsetfillopacity{1.000000}
\node[anchor=base west,inner sep=0pt,outer sep=0pt,color=dialinecolor] at
(17.000000\du,4.200000\du){$\vec{h}^{dec}_{l-1}$};
\pgfsetlinewidth{0.100000\du}
\pgfsetdash{}{0pt}
\pgfsetbuttcap
{
\definecolor{diafillcolor}{rgb}{0.000000, 0.000000, 0.000000}
\pgfsetfillcolor{diafillcolor}
\pgfsetfillopacity{1.000000}
\definecolor{dialinecolor}{rgb}{0.000000, 0.000000, 0.000000}
\pgfsetstrokecolor{dialinecolor}
\pgfsetstrokeopacity{1.000000}
\draw (30.000000\du,0.000000\du)--(28.000000\du,1.000000\du);
}
\pgfsetlinewidth{0.100000\du}
\pgfsetdash{}{0pt}
\pgfsetbuttcap
{
\definecolor{diafillcolor}{rgb}{0.000000, 0.000000, 0.000000}
\pgfsetfillcolor{diafillcolor}
\pgfsetfillopacity{1.000000}
\definecolor{dialinecolor}{rgb}{0.000000, 0.000000, 0.000000}
\pgfsetstrokecolor{dialinecolor}
\pgfsetstrokeopacity{1.000000}
\draw (28.000000\du,-1.000000\du)--(30.000000\du,0.000000\du);
}
\definecolor{dialinecolor}{rgb}{0.000000, 0.000000, 0.000000}
\pgfsetstrokecolor{dialinecolor}
\pgfsetstrokeopacity{1.000000}
\definecolor{diafillcolor}{rgb}{0.000000, 0.000000, 0.000000}
\pgfsetfillcolor{diafillcolor}
\pgfsetfillopacity{1.000000}
\node[anchor=base west,inner sep=0pt,outer sep=0pt,color=dialinecolor] at (31.000000\du,0.000000\du){Attention};
\pgfsetlinewidth{0.100000\du}
\pgfsetdash{}{0pt}
\pgfsetmiterjoin
{\pgfsetcornersarced{\pgfpoint{0.000000\du}{0.000000\du}}\definecolor{diafillcolor}{rgb}{1.000000, 0.000000, 0.000000}
\pgfsetfillcolor{diafillcolor}
\pgfsetfillopacity{1.000000}
\fill (19.000000\du,24.000000\du)--(19.000000\du,27.000000\du)--(25.000000\du,27.000000\du)--(25.000000\du,24.000000\du)--cycle;
}{\pgfsetcornersarced{\pgfpoint{0.000000\du}{0.000000\du}}\definecolor{dialinecolor}{rgb}{0.000000, 0.000000, 0.000000}
\pgfsetstrokecolor{dialinecolor}
\pgfsetstrokeopacity{1.000000}
\draw (19.000000\du,24.000000\du)--(19.000000\du,27.000000\du)--(25.000000\du,27.000000\du)--(25.000000\du,24.000000\du)--cycle;
}
\definecolor{dialinecolor}{rgb}{0.000000, 0.000000, 0.000000}
\pgfsetstrokecolor{dialinecolor}
\pgfsetstrokeopacity{1.000000}
\definecolor{diafillcolor}{rgb}{0.000000, 0.000000, 0.000000}
\pgfsetfillcolor{diafillcolor}
\pgfsetfillopacity{1.000000}
\node[anchor=base,inner sep=0pt, outer sep=0pt,color=dialinecolor] at (22.000000\du,25.295000\du){Small Energy };
\definecolor{dialinecolor}{rgb}{0.000000, 0.000000, 0.000000}
\pgfsetstrokecolor{dialinecolor}
\pgfsetstrokeopacity{1.000000}
\definecolor{diafillcolor}{rgb}{0.000000, 0.000000, 0.000000}
\pgfsetfillcolor{diafillcolor}
\pgfsetfillopacity{1.000000}
\node[anchor=base,inner sep=0pt, outer sep=0pt,color=dialinecolor] at (22.000000\du,26.095000\du){Masking};
\pgfsetlinewidth{0.100000\du}
\pgfsetdash{}{0pt}
\pgfsetmiterjoin
{\pgfsetcornersarced{\pgfpoint{0.000000\du}{0.000000\du}}\definecolor{diafillcolor}{rgb}{0.000000, 1.000000, 0.000000}
\pgfsetfillcolor{diafillcolor}
\pgfsetfillopacity{1.000000}
\fill (19.000000\du,29.000000\du)--(19.000000\du,31.050000\du)--(25.000000\du,31.050000\du)--(25.000000\du,29.000000\du)--cycle;
}{\pgfsetcornersarced{\pgfpoint{0.000000\du}{0.000000\du}}\definecolor{dialinecolor}{rgb}{0.000000, 0.000000, 0.000000}
\pgfsetstrokecolor{dialinecolor}
\pgfsetstrokeopacity{1.000000}
\draw (19.000000\du,29.000000\du)--(19.000000\du,31.050000\du)--(25.000000\du,31.050000\du)--(25.000000\du,29.000000\du)--cycle;
}
\definecolor{dialinecolor}{rgb}{0.000000, 0.000000, 0.000000}
\pgfsetstrokecolor{dialinecolor}
\pgfsetstrokeopacity{1.000000}
\definecolor{diafillcolor}{rgb}{0.000000, 0.000000, 0.000000}
\pgfsetfillcolor{diafillcolor}
\pgfsetfillopacity{1.000000}
\node[anchor=base,inner sep=0pt, outer sep=0pt,color=dialinecolor] at (22.000000\du,30.220000\du){Power Mel};
\pgfsetlinewidth{0.100000\du}
\pgfsetdash{}{0pt}
\pgfsetbuttcap
{
\definecolor{diafillcolor}{rgb}{0.000000, 0.000000, 0.000000}
\pgfsetfillcolor{diafillcolor}
\pgfsetfillopacity{1.000000}
\pgfsetarrowsend{stealth}
\definecolor{dialinecolor}{rgb}{0.000000, 0.000000, 0.000000}
\pgfsetstrokecolor{dialinecolor}
\pgfsetstrokeopacity{1.000000}
\draw (22.000000\du,29.000000\du)--(22.000000\du,27.000000\du);
}
\pgfsetlinewidth{0.100000\du}
\pgfsetdash{}{0pt}
\pgfsetbuttcap
{
\definecolor{diafillcolor}{rgb}{0.000000, 0.000000, 0.000000}
\pgfsetfillcolor{diafillcolor}
\pgfsetfillopacity{1.000000}
\pgfsetarrowsend{stealth}
\definecolor{dialinecolor}{rgb}{0.000000, 0.000000, 0.000000}
\pgfsetstrokecolor{dialinecolor}
\pgfsetstrokeopacity{1.000000}
\draw (22.000000\du,33.000000\du)--(22.000000\du,31.050000\du);
}
\definecolor{dialinecolor}{rgb}{0.000000, 0.000000, 0.000000}
\pgfsetstrokecolor{dialinecolor}
\pgfsetstrokeopacity{1.000000}
\definecolor{diafillcolor}{rgb}{0.000000, 0.000000, 0.000000}
\pgfsetfillcolor{diafillcolor}
\pgfsetfillopacity{1.000000}
\node[anchor=base west,inner sep=0pt,outer sep=0pt,color=dialinecolor] at (22.781899\du,32.109051\du){original speech};
\pgfsetlinewidth{0.100000\du}
\pgfsetdash{}{0pt}
\pgfsetmiterjoin
{\pgfsetcornersarced{\pgfpoint{0.000000\du}{0.000000\du}}\definecolor{diafillcolor}{rgb}{1.000000, 1.000000, 0.000000}
\pgfsetfillcolor{diafillcolor}
\pgfsetfillopacity{1.000000}
\fill (11.000000\du,-18.000000\du)--(11.000000\du,-15.950000\du)--(16.915000\du,-15.950000\du)--(16.915000\du,-18.000000\du)--cycle;
}{\pgfsetcornersarced{\pgfpoint{0.000000\du}{0.000000\du}}\definecolor{dialinecolor}{rgb}{0.000000, 0.000000, 0.000000}
\pgfsetstrokecolor{dialinecolor}
\pgfsetstrokeopacity{1.000000}
\draw (11.000000\du,-18.000000\du)--(11.000000\du,-15.950000\du)--(16.915000\du,-15.950000\du)--(16.915000\du,-18.000000\du)--cycle;
}
\definecolor{dialinecolor}{rgb}{0.000000, 0.000000, 0.000000}
\pgfsetstrokecolor{dialinecolor}
\pgfsetstrokeopacity{1.000000}
\definecolor{diafillcolor}{rgb}{0.000000, 0.000000, 0.000000}
\pgfsetfillcolor{diafillcolor}
\pgfsetfillopacity{1.000000}
\node[anchor=base,inner sep=0pt, outer sep=0pt,color=dialinecolor] at (13.957500\du,-16.780000\du){Decision};
\pgfsetlinewidth{0.100000\du}
\pgfsetdash{}{0pt}
\pgfsetbuttcap
{
\definecolor{diafillcolor}{rgb}{0.000000, 0.000000, 0.000000}
\pgfsetfillcolor{diafillcolor}
\pgfsetfillopacity{1.000000}
\pgfsetarrowsstart{stealth}
\definecolor{dialinecolor}{rgb}{0.000000, 0.000000, 0.000000}
\pgfsetstrokecolor{dialinecolor}
\pgfsetstrokeopacity{1.000000}
\draw (14.000000\du,-20.000000\du)--(13.957500\du,-18.000000\du);
}
\pgfsetlinewidth{0.100000\du}
\pgfsetdash{}{0pt}
\pgfsetmiterjoin
{\pgfsetcornersarced{\pgfpoint{0.000000\du}{0.000000\du}}\definecolor{diafillcolor}{rgb}{1.000000, 1.000000, 0.000000}
\pgfsetfillcolor{diafillcolor}
\pgfsetfillopacity{1.000000}
\fill (2.000000\du,-16.000000\du)--(2.000000\du,-13.950000\du)--(7.915000\du,-13.950000\du)--(7.915000\du,-16.000000\du)--cycle;
}{\pgfsetcornersarced{\pgfpoint{0.000000\du}{0.000000\du}}\definecolor{dialinecolor}{rgb}{0.000000, 0.000000, 0.000000}
\pgfsetstrokecolor{dialinecolor}
\pgfsetstrokeopacity{1.000000}
\draw (2.000000\du,-16.000000\du)--(2.000000\du,-13.950000\du)--(7.915000\du,-13.950000\du)--(7.915000\du,-16.000000\du)--cycle;
}
\definecolor{dialinecolor}{rgb}{0.000000, 0.000000, 0.000000}
\pgfsetstrokecolor{dialinecolor}
\pgfsetstrokeopacity{1.000000}
\definecolor{diafillcolor}{rgb}{0.000000, 0.000000, 0.000000}
\pgfsetfillcolor{diafillcolor}
\pgfsetfillopacity{1.000000}
\node[anchor=base,inner sep=0pt, outer sep=0pt,color=dialinecolor] at (4.957500\du,-14.780000\du){Embedding};
\definecolor{dialinecolor}{rgb}{0.000000, 0.000000, 0.000000}
\pgfsetstrokecolor{dialinecolor}
\pgfsetstrokeopacity{1.000000}
\definecolor{diafillcolor}{rgb}{0.000000, 0.000000, 0.000000}
\pgfsetfillcolor{diafillcolor}
\pgfsetfillopacity{1.000000}
\node[anchor=base west,inner sep=0pt,outer sep=0pt,color=dialinecolor] at
(22.381677\du,28.163576\du){$\vec{x}[m]$};
\end{tikzpicture}